\documentclass{emulateapj}  % for emulate style
\usepackage{epsf}
\usepackage{subfigure}
\usepackage{latexsym}  % for emulate style
\usepackage{epsfig}
\usepackage{natbib}
\usepackage{multirow}

\def\msun{{\rm M}_\odot}
\def\hmpc{\; h \,{\rm Mpc}^{-1}}
\def\hkpc{\; h \,{\rm kpc}^{-1}}
\def\beq{\begin{equation}}
\def\eeq{\end{equation}}

\def\m500{M_{500}}
\def\r500{R_{500}}
\def\rvir{R_{\rm vir}}
\def\mvir{M_{\rm vir}}
\def\hmsun{\; h^{-1} \; M_{\odot}}
\def\los{{\bf \hat n}}
\def\vvec{{\bf v}}

\def\qvec{{\bf \tilde {q}}}
\def\qvecreal{{\bf {q}}}
\def\kvec{{\bf  k}}
\def\Pdd{P_{\delta \delta}}
\def\Pvv{P_{vv}}
\def\Pdv{P_{\delta v}}
\def\kvecunit{{\bf \hat k}}
\def\Tcmb{T_{\rm cmb}}

\def\kjeans{k_{\rm J}}

\def\del{{\bf {\bigtriangledown}}}
\def\mksq{{\mu {\rm K^2}}}
\def\drm{{d}}
\def\zrei{z_{\rm rei}}
\def\dell{\mathcal{D}_\ell}
\def\dthree{\mathcal{D}_{3000}}

\begin{document}

\bibliographystyle{apj} 
%\bibstyle{aas}

\shorttitle{The kinetic SZ power spectrum}
\shortauthors{Shaw et al.}

\author{Laurie D. Shaw\altaffilmark{1,2}}
\author{Douglas H. Rudd\altaffilmark{1,2}}
\author{Daisuke Nagai\altaffilmark{1,2}}
\altaffiltext{1}{Department of Physics, Yale University, New Haven, CT 06520}
\altaffiltext{2}{Yale Center for Astronomy \& Astrophysics, Yale University, New Haven, CT 06520}
\email{laurie.shaw@yale.edu}

\title{Deconstructing the kinetic SZ Power Spectrum} 
\begin{abstract}
We present a detailed investigation of the impact of astrophysical
processes on the shape and amplitude of the kinetic SZ (kSZ) power
spectrum from the post-reionization epoch. This is achieved by
constructing a new model of the kSZ power spectrum which we calibrate
to the results of hydrodynamic simulations. By construction, our
method accounts for all relevant density and velocity modes and so is
unaffected by the limited box size of our simulations. We find that
radiative cooling and star-formation can reduce the amplitude of the
kSZ power spectrum by up to $33 \%$, or $1 \mksq$ at $\ell =
3000$. This is driven by a decrease in the mean gas density in groups
and clusters due to the conversion of gas into stars. Variations in
the redshifts at which helium reionization occurs can effect the
amplitude by a similar fraction, while current constraints on
cosmological parameters (namely $\sigma_8$) translate to a further
$\pm 15 \%$ uncertainty on the kSZ power spectrum.  We demonstrate how
the models presented in this work can be constrained -- reducing the
astrophysical uncertainty on the kSZ signal -- by measuring the
redshift dependence of the signal via kSZ tomography. Finally, we
discuss how the results of this work can help constrain the duration
of reionization via measurements of the kinetic SZ signal sourced
by inhomogeneous (or patchy) reionization.


\end{abstract}

\section{INTRODUCTION}

In the last few years, significant progress has been made towards making
precision measurements of the temperature anisotropy in the cosmic
microwave background (CMB) at sub-degree scales.  Ground-based
experiments such as the South Pole Telescope \citep{ruhl04} and the
Atacama Cosmology Telescope \citep{kosowsky03} have achieved the
necessary combination of sensitivity and angular resolution to measure
the primary CMB temperature anisotropy power spectrum to the seventh
acoustic peak and beyond \citep{dunkley10, keisler11}. The
Planck\footnote{http://www.rssd.esa.int/index.php?project=Planck}
satellite will measure the CMB temperature power spectrum at
cosmic-variance limited precision to angular scales below one tenth of
a degree.

On angular scales smaller than $\sim$4 arcminutes ($\ell \gtrsim 2700$),
the CMB power spectrum is dominated by `secondary' anisotropies,
temperature fluctuations that are generated by the interaction of CMB
photons with large scale structure between the observer and the
surface of last scattering. The principal contribution to the
secondary anisotropy signals comes from the Sunyaev-Zel'dovich (SZ)
effect, which can be broken down into two components-- the `thermal'
and `kinetic' effects. The former describes the transfer of thermal
energy from free electrons in the hot intra-cluster medium to CMB
photons via inverse-Compton scattering. CMB photons receive a boost in
energy, distorting the Planckian form of its spectrum. This gives the
thermal SZ (tSZ) effect a unique frequency dependence which can be
utilized by experiments to extract this signal from the primary
CMB signal and foregrounds. The kinetic SZ (kSZ) effect is caused by
the doppler shifting of CMB photons via scattering off clouds of
electrons with a non-zero bulk velocity (along the line-of-sight)
relative to the CMB rest frame. Unlike the tSZ effect, the kSZ has the
same frequency dependence as the primary CMB.

The thermal SZ effect has now been detected for both large numbers of
individual clusters \citep{vanderlinde10, marriage10, williamson11,
  planckesz, marrone11} and as a secondary anisotropy signal in the
CMB power spectrum \citep{lueker10,dunkley10, shirokoff11}. While the
kSZ effect has not yet been detected in either case, it is likely be
done so first in the power spectrum. CMB temperature fluctuations
sourced by the kSZ effect are proportional to the product of the
electron density and line-of-sight velocity. The lack of an electron
temperature weighting means that the contribution of low-temperature
gas is more significant than for the tSZ effect (which is proportional
to the product of electron density and temperature). On the other
hand, the kSZ signal from individual groups and clusters is weaker.

The kSZ power spectrum is also sensitive to the details of
reionization \citep{gruzinov98, knox98, mcquinn05, zahn05,
  iliev07}. In models of {\em inhomogeneous} or {\em patchy}
reionization -- in which different regions of the Universe are
reionized at different times -- bubbles of free electrons around
UV-emitting sources are embedded in an otherwise neutral medium. If
these bubbles have a large-scale bulk velocity then they will impart a
temperature anisotropy onto the CMB. In this work we refer to the kSZ
effect from reionization as the `patchy' signal, while that from
epochs after reionization is complete is referred to as the
`post-reionization' signal.

\citet{zahn05} and \citet{mcquinn05} demonstrated that, to first
order, the magnitude of the kSZ power from patchy reionization is
dependent on the duration of reionization.  Hence, when combined with
measurements of the optical depth to reionization from the primary CMB
power spectrum, the redshift range spanned by the epoch of
reionization can be constrained. For example, if reionization started
at $z = 14$ and ended at $z = 6$, it would generate roughly $3 \mksq$
of patchy kSZ power (at $\ell = 3000$), while the range $8 \leq z \leq
12$ would generate $1.5 \mksq$ \citep{mcquinn05}. While there are
several methods for probing the redshift at which reionization ended
\citep[e.g.,][and references therein]{oh05, fan06, lidz06, becker07,bolton07, ouchi10,
  mortlock11}, there are currently no other means by which the
duration of reionization can be measured, making the kSZ power
spectrum a unique and exciting probe.

The patchy kSZ signal is expected to have a slightly different angular
shape to the post-reionization signal, peaking at a larger angular
scale \citep[$\ell \approx 2000$,][]{zahn05, zahn11}, corresponding
to the characteristic bubble size during reionization. In principle,
the two components of the kSZ signal can be separated by a precise measurement
of the power spectrum that encompasses a wide range of angular
scales. In practice, however, the primary CMB signal swamps that of
the kSZ at $\ell < 3000$, while extra-galactic foregrounds dominate at
$\ell > 5000$. Hence, differentiating the post-reionization kSZ power
from that sourced by patchy reionization via their angular scale
dependence is an intractable task. To measure the patchy component of
this signal it is therefore vitally important to have a good
theoretical understanding of the post-reionization contribution.

One of the principal aims of this work is to investigate the
theoretical uncertainty in the post-reionization kSZ power
spectrum. Specifically, we aim to investigate the impact of
astrophysical processes such as radiative cooling of gas, the
formation of stars and galaxies and feedback from supernovae. We
construct an analytic model for the kSZ power spectrum which we
calibrate to the results of hydrodynamic simulations. These
simulations are run in both the non-radiative regime, and including
radiative cooling and star-formation. We are thus able to compare the
impact of these processes on the kSZ power spectrum.  
Our model also enables us to investigate the cosmological
scaling of the kSZ power spectrum.

Several previous studies have measured the kSZ power spectrum directly
from cosmological simulations by generating synthetic sky maps,
projecting through the simulation box stacked over multiple timesteps
\citep{white02, hallman09, battaglia10}. The principal drawback to
this approach is that very large simulation box-sizes are necessary to
adequately capture the large-scale velocity flows that contribute
significantly to the kSZ signal at small angular scales
\citep{zhang04}. As we shall demonstrate, simulations with box sizes
significantly less than 1 Gpc/$h$, systematically and substantially
underestimate the kSZ power spectrum. Furthermore, high spatial
resolution is required to resolve small-scale baryonic processes such
as cooling and star-formation.

In this work we adopt a hybrid approach. We use high-resolution
simulations to capture the effect of non-linear structure formation on
the gas density power spectrum, and use the results to improve our
analytic calculation for the kSZ effect. Our method accounts for all
relevant velocity and density modes, and circumvents the prohibitive
requirement of high spatial resolution and extremely large
cosmological boxes.

This paper is organized as follows: in Section \ref{sec:model} we
provide an overview of the kSZ effect, describe our model for the kSZ
power spectrum, and discuss how we implement the modifications
required to account for the astrophysical processes in our
simulations. In Section \ref{sec:simulations} we describe the
hydrodynamic simulations used to calibrate our model. In Section
\ref{sec:results} we present results from the simulations, focusing
specifically on the gas density and momentum power spectra and
comparing these with the predictions of our models. In Section
\ref{sec:ksz} we discuss the kSZ power spectra predicted by our
models, their scaling with cosmological parameters, and the means by
which they could be distinguished observationally. We also investigate
the impact of helium reionization on the power spectrum. Finally, we
compare our model with the results of previous work and the latest
observations.

Except when referring to specific simulations with specific
cosmological parameters, throughout this paper we assume a fiducial,
spatially-flat, $\Lambda$CDM cosmological model consistent with the
WMAP7 best-fit cosmological parameters, namely $H_0 = 71$ km
s$^{-1}$Mpc$^{-1}$, $\Omega_M = 0.27$, $\Omega_b = 0.047$,
$\Omega_\Lambda = 0.73$, $n_s = 0.95$ and $\sigma_8 = 0.82$.

\section{Modeling the kinetic SZ Power Spectrum}
\label{sec:model}
\subsection{kSZ Basics}
\label{sec:kszbasics}
Thomson scattering of CMB photons off clouds of free electrons with a
coherent bulk velocity along the line-of-sight from the observer
produce fluctuations in the observed brightness temperature of the CMB,
\beq
\frac{\Delta T}{\Tcmb}(\los) = \frac{\sigma_T}{c} \int_0^{z_{\rm rei}} \frac{\drm x}{\drm z}\frac{\drm z}{(1+z)} \exp(-\tau(z)) n_e(z)  \vvec\cdot\los \;,
\eeq
where $\sigma_T$ is the Thomson cross-section for an electron, $x$ is
the comoving distance to redshift $z$, $n_e$ is the free electron
number density and $\vvec\cdot\los$ is the component of the peculiar
velocity of the electrons along the line-of-sight. We are principally
concerned with the kinetic SZ power spectrum in the post-reionization
era, so the upper limit to the integral, $z_{\rm rei}$, corresponds to
the redshift at which reionization ends (i.e., hydrogen has been
entirely ionized). Unless stated otherwise we assume a fiducial value
of $z_{\rm rei} = 10$.

$\tau$ is the Thomson optical depth,
\beq 
\tau(z) = \sigma_T \int_0^{z} \frac{\bar{n}_e(z')}{1+z'} \frac{\drm x}{\drm z'} \drm z' \;,  
\label{eq:tau}
\eeq
where $\bar{n}_e$ is the mean free-electron density,
\beq
\bar{n}_e = \frac{\chi \rho_{\rm g}(z)}{\mu_e m_p} \;\;,
\label{eq:ne}
\eeq 
$\rho_{\rm g}(z) = \rho_{\rm g,0}(1+z)^3$ is the mean gas density
density of the Universe at redshift $z$, and $\mu_e m_p$ is the mean
mass per electron, where $\mu_e = 1.14$. We define $\chi$ as
the fraction of the total number of electrons that are ionized. We
assume that, for $z < z_{\rm rei}$, hydrogen is completely ionized, so
$\chi$ is dependent on the abundance and ionization state of helium,
\beq 
\chi = \frac{1 - Y_p(1 - N_{\rm He}/4)}{1-Y_p/2} \;\;,
\label{eq:chi}
\eeq
where $Y_p$ is the primordial helium abundance and $N_{\rm He}$ the
number of helium electrons ionized (which can be a function of
redshift). Thus for $Y_p = 0.24$, $\chi = $\{0.86, 0.93, 1\} for
neutral, singly, and fully ionized helium. For our fiducial model we
assume $\chi = 0.86$ (i.e., $N_{\rm He} = 0$) at all redshifts. We
explore the effect of helium reionization on the kSZ power spectrum in
Section \ref{sec:ksz}.

Writing $n_e = \bar{n}_e(1+\delta)$ we define the density weighted
peculiar velocity $\qvecreal = \vvec (1+\delta)$, so that
\beq 
\frac{\Delta T}{\Tcmb}(\los) = \frac{\sigma_T \rho_{\rm g,0}}{\mu_e m_p c} \int_0^{z_{\rm rei}}  \frac{\drm x}{\drm z} \drm z (1+z)^2  \chi \exp(-\tau(z)) \los\cdot\qvecreal \;.
\label{eq:ksz}
\eeq 

KSZ temperature fluctuations are generated by the projected
contribution of ionized gas with a non-zero peculiar velocity along
the line of sight. A key property of the kSZ signal is that Fourier
modes of $\qvecreal$ ($\qvec(\kvec)$) which have $\kvec$ parallel to
$\los$ suffer severe cancellation when projected along the line of
sight. Therefore, only modes of $\qvec(\kvec)$ parallel to $\los$ but
perpendicular to $\kvec$ can contribute \citep[see][for a rigorous
  demonstration of this. Henceforth we use $\tilde{}$ to denote a
  Fourier space quantity.]{jaffe98}.

$\qvecreal$ can be decomposed into divergence free
($\qvecreal_B$) and curl-free ($\qvecreal_E$) components, which
satisfy $\del . \qvecreal_B = 0$ and $\del \times \qvecreal_E = 0$,
respectively. In the Fourier domain, $\qvec = \qvec_E + \qvec_B$,
where $\qvec_E = \kvecunit (\qvec \cdot \kvecunit)$ and
\beq
\qvec_{B}(\kvec) = \qvec - \kvecunit ( \qvec \cdot \kvecunit) \;.
\eeq
When projected along the line of sight, the peaks and troughs of the
component of $\qvec$ parallel to $\kvec$, $\qvec_E$, will
cancel. Therefore, only the component of $\qvec$ perpendicular to
$\kvec$, $\qvec_B$, contributes to the kSZ signal
\citep{vishniac87,jaffe98}.

$\qvec_B$ can be written as a convolution between the Fourier transform of the velocity and density fields,
\beq
\qvec_{B}(\kvec) = \int \frac{d^3 \kvec'}{(2\pi)^3}(\kvecunit' - \mu \kvecunit)\tilde{v}(k')\delta_b(|\kvec - \kvec'|) \;\;,
\label{eq:qvec}
\eeq 
where $\mu = \kvecunit \cdot \kvecunit'$. In the linear regime, ${\bf
  \tilde{v}(k)}$ is parallel to $\kvecunit$, so only the $\vvec
\delta$ component of $\qvecreal$ can contribute to
$\qvecreal_{B}$. The kSZ power spectrum is thus generated by the
cross-term $\delta \vvec$ \citep{ostriker86, vishniac87}.

In the small angle limit, the kSZ angular power spectrum can be calculated using Limber's approximation,
\begin{eqnarray}
C_\ell & = & \frac{8 \pi^2}{(2\ell+1)^3} \left(\frac{\sigma_T \rho_{\rm g,0}}{\mu_e m_p c}\right)^2 \int_0^{z_{rei}} (1+z)^4 \chi^2 \Delta_B^2(\ell/x,z) \nonumber\\
& &  \times \exp\bigl(-2 \tau(z)\bigr) x \frac{\drm x}{\drm z} \drm z \;\; 
\label{eq:cl}
\end{eqnarray}
where $k = \ell / x$, $\Delta_B^2(k,z) = k^3 P_B(k,z)/(2\pi^2)$, and the power
spectrum of the curl component of the momentum field, $P_B$, is
defined by $(2\pi)^3P_B \delta(\kvec-\kvec')= \langle \qvec_B(\kvec)
\qvec_B^*(\kvec')\rangle$.

\citet{vishniac87} first calculated an expression for $\Delta_B^2$
\citep[see also][]{jaffe98, dodelson95, ma02},
\begin{eqnarray}
\Delta^2_B(k) & = & \frac{k^3}{2\pi^2}\int \frac{d^3 k'}{(2\pi)^3} \Bigl[ (1-\mu^2)\Pdd(|\kvec-\kvec'|)P_{vv}(k') \Bigr. \nonumber\\
& & \Bigl. - \frac{(1-\mu^2)k'}{|\kvec-\kvec'|}P_{\delta v}(|\kvec-\kvec'|)P_{\delta v}(k')\Bigr] \;\;,
\label{eq:Pq}
\end{eqnarray}
where $\Pdd$ and $\Pvv$ are the linear theory density and velocity
power spectra, and $\Pdv$ is the density-velocity cross spectrum (for
clarity we have suppressed the redshift dependence of $\Delta^2_B(k)$
and $P(k)$).  In the linear regime, the continuity equation allows us
to relate the peculiar velocity field with density perturbations,
$\tilde{\vvec}(\kvec) = i \kvecunit (f \dot{a}/k)
\tilde{\delta}(\kvec)$, where $f = \drm \log{D}/ \drm \log{a}$ and $D$
is the linear growth factor. Therefore,

\beq
P_{vv}(k) = \left(\frac{f \dot{a}}{k}\right)^2 P_{\delta \delta}(k) \;;\;\; P_{\delta v}(k) = \left(\frac{f \dot{a}}{k}\right) P_{\delta \delta}(k) \;\;.
\eeq
Plugging these into Equation \ref{eq:Pq}, we obtain
\beq
\Delta_B^2(k) = \frac{k^3}{2 \pi^2} \dot{a}^2 f^2 \int \frac{d^3 k'}{(2 \pi)^3} \Pdd(|\kvec-\kvec'|)\Pdd(k') I(k,k') \;\;,
\label{eq:deltab_lin}
\eeq
where
\begin{equation}
I(k,k') = \frac{k(k-2k'\mu)(1-\mu^2)}{k'^2(k^2 + k'^2 - 2kk'\mu)} \;.\nonumber
\end{equation}

Combining Equations \ref{eq:cl} and \ref{eq:Pq} gives the well-known
Ostriker-Vishniac effect, which is the linear-theory part of the kinetic SZ
power spectrum.

\subsection{Non-linear Contributions}
\label{sec:nonlinear_corrections}

Several previous studies have investigated the impact of non-linear
structure formation on the kSZ power spectrum, showing that non-linear
corrections become large for angular scales $\ell > 1000$
\citep{hu00,ma02,zhang04}. \citet{hu00} and \citet{ma02} demonstrate
that one can include the effect of non-linear density fluctuations to
the kSZ power spectrum by exchanging the linear theory matter power
spectrum, $\Pdd$, in Equation \ref{eq:deltab_lin} with the non-linear matter
power spectrum, $\Pdd^{\rm NL}$, so that

\beq
\Delta_B^2(k) = \frac{k^3}{2 \pi^2} \dot{a}^2 f^2 \int \frac{d^3 k'}{(2 \pi)^3} \Pdd^{\rm NL}(|\kvec-\kvec'|)\Pdd(k') I(k,k') \;\;.
\label{eq:deltab_seminl}
\eeq \citet{ma02} argue that the kinetic SZ signal is less sensitive
to non-linear velocity fluctuations than non-linear density
fluctuations due to the $1/k^2$ weighting in the former. Using
hydrodynamical simulations, \citet{zhang04, shao11} show that, for $k
> 2 \hmpc$, the power in the curl component of the velocity field
(generated by non-linear gravitational collapse) can exceed the linear
theory prediction. To account for this,
they suggest an phenomenological correction in which $\Pdd(k')$ is
replaced by its non-linear counterpart. In this work, we follow the
non-linear correction proposed by \citet{ma02} given in Equation
\ref{eq:deltab_seminl}. However, we investigate to what degree
velocity fluctuations in the non-linear regime may effect the kSZ
power spectrum in Section \ref{sec:ksz}.

Throughout this work, we calculate the non-linear density power
spectrum, $\Pdd^{\rm NL}$ using the {\sc HaloFit} prescription of
\citet{smith03}. We find that the dark matter power spectrum predicted
by {\sc HaloFit} is within 8\% of that measured in our non-radiative
simulations for $k < 4 \hmpc$ \citep{rudd08}, although it
systematically underestimates the power spectrum at $k > 1
\hmpc$. \citet{heitmann10} found similar results comparing {\sc
  HaloFit} to a large suite of N-body simulations over a wide range of
cosmological models.

There is a much larger discrepancy between {\sc HaloFit} and our
simulation that includes cooling and star-formation (CSF). The
formation of dense clumps of stars and gas at the center of halos
produces a steepening of the dark matter density profile at small
radii, increasing the density of the halo core. This process is
sometimes referred to as  `halo (or adiabatic)  contraction'
\citep{blumenthal86, gnedin04, gnedin11}. The matter power spectrum measured in
dark matter-only simulations, or by halo models calibrated on such
simulations (such as {\sc HaloFit}), no longer matches that measured
in simulations that include radiative cooling at wavenumbers $k > 1
\hmpc$ \citep{jing06, rudd08, duffy10, vanDaalen11}.

We incorporate the effects of halo contraction using the simple
modification to {\sc HaloFit} suggested by \citet{rudd08}. This is
implemented by multiplying the matter power spectrum by the ratio of
Fourier-transformed NFW density profiles \citep{navarro97} with two
different concentrations,
\beq 
P_{\rm DMcsf}(k,z) = \Phi(k,z)^2 P_{\rm DMonly}(k,z)\;,
\label{eq:adiabatic_contraction}
\eeq
where 
\beq
\Phi(k,z) = \left[\frac{\lambda(R_{\rm vir}k/c_2,c_2)}{\lambda(R_{\rm vir}k/c_1,c_1)}\right] \;.
\eeq
$\lambda(R_{\rm vir},k,c)$ is the Fourier transform of an NFW halo of
virial radius $R_{\rm vir}$\footnote{The virial mass and radius are
  defined by $\mvir = \frac{4}{3} \pi \rvir^3 \Delta_c \rho_c(z)$,
  where $\Delta_c$ is the virial overdensity given in \citet{bryan98}
  and $\rho_c(z)$ is the critical density at redshift $z$}, and
($c_2$, $c_1$) are the concentrations of the halo including and
omitting the effects of radiative cooling and star-formation,
respectively\footnote{See \citet[]{rudd08} for an analytic expression
  for $\lambda$.}.  We apply the correction to the matter power
spectrum given by Equation \ref{eq:adiabatic_contraction}, using $c_1
= 5$, $c_2 = 8.5$ and $R_{\rm vir}(z = 0.55) = 1.1$ comoving Mpc/$h$
(corresponding to a virial mass of $M_{\rm vir} = 1.07 \times 10^{14}
\hmsun$).

This procedure is clearly a very simplistic approximation to the
effects of halo contraction on the matter power spectrum. The
correction to the power spectrum is determined entirely by the effect
of a change in concentration on the density profile of a halo with a
characteristic mass and redshift. However, in Section
\ref{sec:results} we demonstrate that it does provide a good agreement
between our model and the CSF simulations at small scales.

\begin{figure}
%\plotone{f1.eps}
\includegraphics[scale = 0.30]{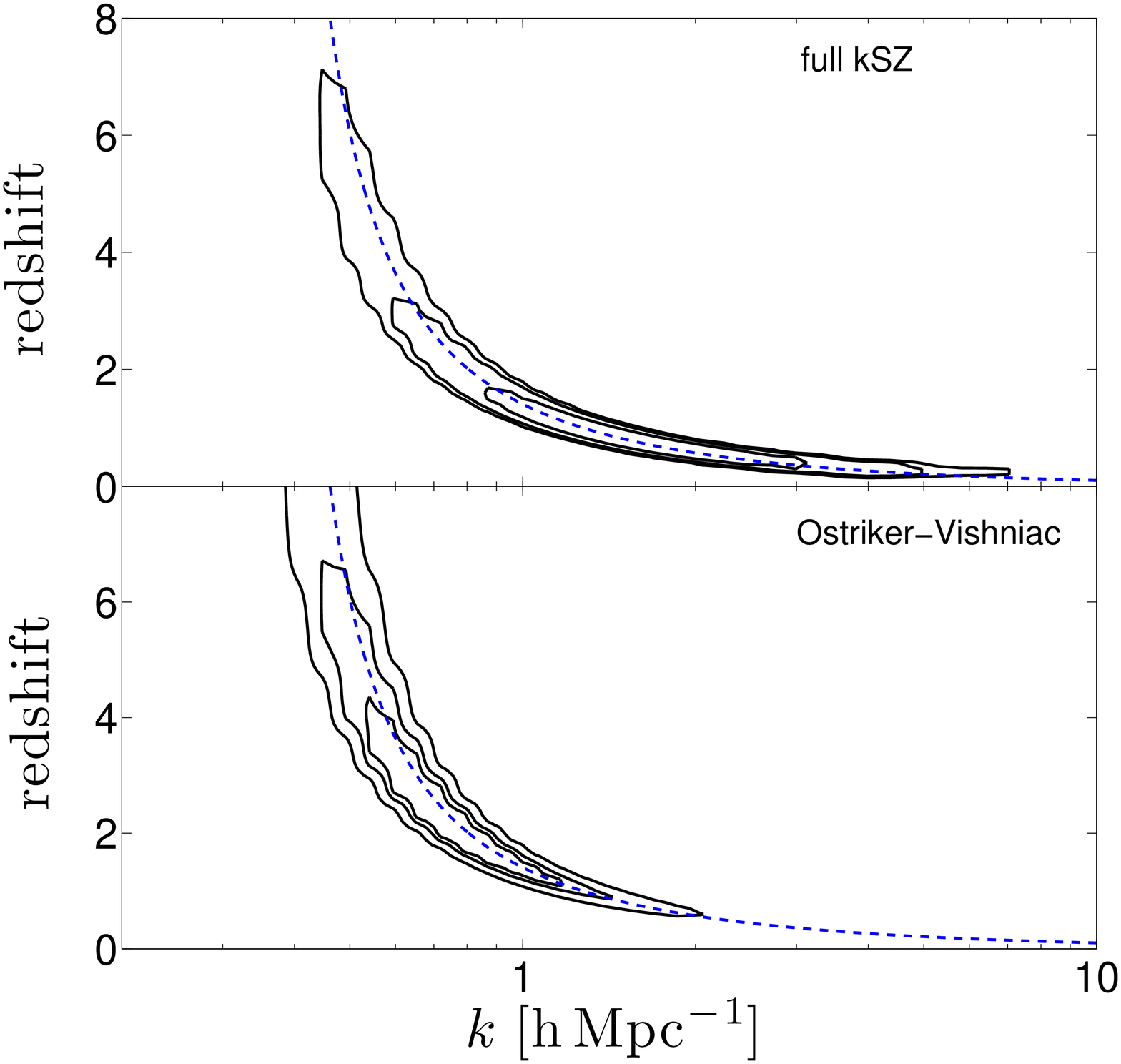}
\caption{The contribution, $\drm^2 C_\ell / \drm z \drm \ln(k)$, to
  the total kinetic SZ power at $\ell = 3000$ by density fluctuations
  at a given redshift and spatial scale. In each panel, the contours
  enclose the regions that contribute 25, 50 and 75 \% of the total
  power (from inner to outer contours). The top panel shows mass and
  redshift dependence for the full kinetic SZ power spectrum including
  the non-linear density fluctuations. The lower panel shows the same
  contours for the Ostriker-Vishniac effect, in which only the linear
  density power spectrum is used. The dashed blue lines represent the
  size-distance relation, $k = \ell / x(z)$, for $\ell =
  3000$. Kinetic SZ power is generated by density fluctuations in the
  range $0.3 \leq k \leq 10 \hmpc$, with non-linear fluctuations
  boosting the signal in the high-$k$, low redshift regime.}
\label{fig:ksz_contours}
\end{figure}

\subsection{Thermal Pressure of Baryons}

In previous work, it has commonly been assumed that $\delta_{\rm gas} =
\delta_{\rm DM}$, that is, perturbations in the gas density exactly
follow those of the dark matter at all scales \citep{dodelson95,
  jaffe98, ma02, zhang04}. However, thermal pressure between baryons
should erase density fluctuations in the gas distribution at small
scales. Furthermore, radiative cooling and star-formation can
significantly modify the gas density power spectrum at both large and
small scales \citep{jing06, rudd08, casarini11, vanDaalen11}.

In order to accurately predict the kSZ power spectrum we must be able to relate
the power spectrum of gas density fluctuations, $P_{\rm gas}$ (or more precisely,
fluctuations in the number density of free electrons), with that of the dark 
matter. We incorporate the effects of baryon physics into a 
window function $W(k)$, such that 
\beq 
P^{\rm NL}_{\rm gas}(k,z) = W^2(k,z) P^{\rm NL}_{\rm DM}(k,z)\;. 
\eeq 
This then replaces the non-linear matter power spectrum in Equation
\ref{eq:deltab_seminl}. Note that we assume the velocity field of the
gas follows that of the dark matter, which is a reasonable assumption
at large scales.

In the non-radiative regime, the qualitative shape of $W^2(k)$ is simple
to imagine. At large scales, and before the onset of gravitational
collapse, we expect $W^2(k) \approx 1$. However, at small scales, the gas
thermal pressure force suppresses gas density perturbations and so
$W^2(k)$ will tend to zero as $k$ increases. Therefore, $W^2(k)$ acts as
a filter, smoothing the gas density at some characteristic scale. As
we shall demonstrate, the form of $W^2(k)$ for our simulation including
radiative cooling and star-formation has a more complex dependence on $k$.

\citet{gnedin98} demonstrate that, for coupled density perturbations
in the linear regime, the ratio of gas to dark matter density
fluctuations is well described by the form 
\beq
W_{\rm G98}(k,z) = 0.5\left(\exp{(-k^2/k_f^2)} + \frac{1}{[1 + 4(k/k_f)^2]^{1/4}}\right) \;,
\label{eq:wk_lin}
\eeq
where the (redshift-dependent) characteristic filter scale $k_f$ is
given by
\beq 
\frac{1}{k_f^2(t)} = \frac{1}{D(t)}\int_0^t a^2(t') \drm t'\frac{\ddot{D}(t') +
  2H(t')\dot{D}(t')}{k_J^2(t')} \int_{t'}^t\frac{\drm t''}{a^2(t'')} \;. 
\label{eq:kf}
\eeq 
Here $\kjeans$ is the Jean's scale, 
\beq 
\kjeans(t) = \frac{a}{c_{\rm S}(t)}\sqrt{4\pi G \rho_m(t)} \;, 
\eeq 
$\rho_m$ is the mean matter density of the Universe and $c_{\rm S} =
\sqrt{d P/d \rho}$ is the mean sound speed at time $t$. As
noted by \citet{gnedin03}, the filter scale at a given time is not
directly proportional to the Jeans scale at the same time, but to the
integral over the thermal history of the gas up to that point.

\citet{hu00} and \citet{mcquinn05} used Eq.~\ref{eq:wk_lin} to
approximate the effects of thermal pressure in their calculations of
the (post-reionization) kSZ power spectrum. In this work, we measure
the shape and evolution of $W^2(k)$ in hydrodynamic simulations
and apply the results to our model for the power spectrum.

\subsection{Final Model}
\label{sec:final_model}

Our full model for the kSZ power spectrum is given by Equation
\ref{eq:cl}, where the expression for $\Delta_B^2$ including
non-linear corrections, the gas window function and halo
contraction is
\begin{eqnarray} 
\Delta_B^2(k) & = & \frac{k^3}{2 \pi^2} \dot{a}^2 f^2 \int \frac{d^3 k'}{(2 \pi)^3} W^2(|\kvec-\kvec'|) \Phi^2(|\kvec-\kvec'|) \nonumber \\
& & \times \Pdd^{\rm NL}(|\kvec-\kvec'|) \Pdd(k') I(k,k') \;\;.
\label{eq:deltab_ksz}
\end{eqnarray}
We remind the reader that the superscript NL represents the non-linear
matter power spectrum, calculated using the {\sc HaloFit} procedure of
\citet{smith03}. Otherwise, we use the linear-theory matter power
spectrum.

One of the principal aims of this work is to explore the impact of
astrophysical processes such as radiative cooling, star-formation and
supernova feedback on the kSZ power spectrum. This is achieved by
measuring the window function $W^2(k)$ in hydrodynamic simulations
both including and omitting these processes. By plugging the results
into our model, we are able to calculate the effect of these processes
on the kSZ power spectrum. To investigate the impact of non-linear
structure formation and halo contraction, we measure the curl
component of the momentum power spectrum, $\Delta^2_B$, directly from
our simulations and compare with the analytic prediction given by
Equation \ref{eq:deltab_ksz}.

Based on our hydrodynamic simulations, we investigate three models
for the kinetic SZ power spectrum, labeled DM (dark matter), NR
(non-radiative) and CSF (cooling and star-formation). The three models
differ principally in the gas window function, $W^2_i(k)$, used to relate
the non-linear dark matter power spectrum to the gas density power
spectrum, where $i = \{1, W^2_{\rm NR}, W^2_{\rm CSF} \}$ for the
${\rm \{ DM, NR, CSF \}}$ models. For the CSF model we also include
the halo contraction correction to the dark matter density power
spectrum, $\Phi(k,z)$, given by Equation
\ref{eq:adiabatic_contraction}. For the DM and NR models we set this
equal to 1 at all $k$ and $z$.

In our model, simulations are used to determine the gas density power
spectrum whereas the velocity modes are entirely calculated from
linear theory. Our technique thus circumvents the problems relating to
the truncation of large-scale velocity modes due to a limited
simulation box size (see Section \ref{sec:momentum}). Indeed, our
simulations accurately resolve the range of spatial scales over which
density fluctuations contribute significantly to the kinetic SZ power
spectrum for $1000 \leq \ell \leq 10000$.  We now demonstrate this
point explicitly.

In Figure \ref{fig:ksz_contours} we plot the contribution to the total
kinetic SZ power at $\ell = 3000$ by density fluctuations at a given
redshift and spatial scale $k = 3000 / x$. In each panel, the contours
enclose the regions that contribute 25, 50 and 75\% of the total power
(inner to outer contours). The top panel shows contours for the full
kinetic SZ power spectrum including the non-linear density
fluctuations. The lower panel shows the same contours for the
Ostriker-Vishniac effect, in which only the linear-regime matter power
spectrum is used. The dashed blue lines show the relation $k = \ell /
x(z)$ at $\ell = 3000$.  The shape of the contours follow that
expected for the distance-size relation at a constant angular
scale. In this figure $W^2(k)$ is fixed to 1 at all scales.

Comparing the two panels clearly demonstrates the impact of non-linear
density perturbations on the kSZ power. The full kSZ contours extend
to larger $k$ and lower redshift, where the impact of the non-linear
corrections to the power spectrum are largest. More than 50\% of the
full kSZ power is sourced by density fluctuations in the range $0.6
\leq k \leq 5 \hmpc$ and $0 \leq z \leq 3$, while wavenumbers up to $k
= 8\hmpc$ provide a non-negligible contribution.  In contrast, we
find that the principal contribution of {\em velocity} modes to the
kSZ signal comes from the range of scales encompassed by $0.005 \leq k
\leq 0.5 \hmpc$ and thus from much smaller $k$ than the density
fluctuations. This emphasizes that the kinetic SZ power spectrum is
generated by small scale density fluctuations caught up in large scale
velocity flows. The OV signal is spread over a larger redshift range,
but limited to contributions from density fluctuations at $k \leq 2
\hmpc$. Our simulations accurately resolve density fluctuations over
the full range of $k$ plotted and are thus well-suited to calibrating
our model.

\section{Simulations}
\label{sec:simulations}
\begin{deluxetable*}{lcccccccccc}
\tablecolumns{11}
\tablewidth{\linewidth} 
\tablecaption{List of Simulations\label{table:simulations}}
\tablehead{
\multirow{2}{*}{Name\hspace{0.05\linewidth}} &
\multicolumn{1}{c}{$L_{\mathrm{box}}$} &
\multirow{2}{*}{$\Omega_m$} &
\multirow{2}{*}{$\Omega_\Lambda$} &
\multirow{2}{*}{$h$} &
\multirow{2}{*}{$\Omega_b h^2$} &
\multirow{2}{*}{$\sigma_8$} &
\multirow{2}{*}{$n_s$} &
\multirow{2}{*}{$n_{DM}$} &
\multicolumn{1}{c}{$m_p$} &
\multicolumn{1}{c}{$\Delta x$} 
\\
& $(h^{-1} \mathrm{Mpc})$ & & & & & & & & $(10^9 h^{-1} M_\odot)$ & $(h^{-1} \mathrm{kpc})$ }
\startdata
L60NR & 60 & 0.3 & 0.7 & 0.7 & 0.021 & 0.9 & 1.0 & $256^3$ & 0.92 & 1.8 \\
L60CSF & 60 & 0.3 & 0.7 & 0.7 & 0.021 & 0.9 & 1.0 & $256^3$ & 0.92 & 3.6 \\
%CLS & 240 & 0.25 & 0.75 & 0.73 & 0.0224 & 0.8 & 0.95 & $512^3$ & 5.95 & 29 \\
BolshoiNR & 250 & 0.27 & 0.73 & 0.7 & 0.023 & 0.82 & 0.95 & $1024^3$ & 1.08 & 3.8 \\
\vspace*{-0.5em} \enddata \tablecomments{$L_{\mathrm{box}}$ is the
  simulation box side-length, $n_{DM}$ and $m_p$ are the number of
  dark of dark matter particles and their mass (in units of $10^9 \,
  h^{-1} M_\odot$), and $\Delta x$ is the peak spatial resolution of
  the simulation. }
\end{deluxetable*}

Our simulations are performed using the Adaptive Refinement Tree (ART)
$N$-body$+$gas-dynamics code \citep{kravtsov02,rudd08}, which is an
Eulerian code that uses adaptive refinement in space and time, and
non-adaptive refinement in mass \citep{klypin01} to achieve the
dynamic range necessary to resolve the cores of halos formed in
self-consistent cosmological simulations.

The simulation properties, their associated box sizes, cosmological
parameters, and resolutions are summarized in
Table~\ref{table:simulations}.  Two of the three simulations
(BolshoiNR and L60NR) are performed in the non-radiative regime where
baryons are shock heated during structure formation but unable to
radiatively cool.  BolshoiNR is a large simulation designed to study
the formation of galaxy clusters. L60NR and L60CSF are realizations of
the same set of initial density fluctuations simulated with varying
physical processes included and were previously used to study baryon
effects on the matter power spectrum in \citet{rudd08}.

A complete description of the physical processes implemented in
the L60CSF simulation can be found in \citet{rudd08}.  Briefly, 
gas is converted to collisionless stellar particles using a Kennicutt-Schmidt 
type density relation for a gas consumption timescale of $\sim2.7 \mathrm{Gyr}$.
Mass, energy, and metals are returned to the ISM through prescriptions
for types II and Ia supernovae and stellar mass-loss via winds. 
Radiative cooling rates are tabulated over a range of gas temperature,
density, and metallicity including a redshift-dependent cosmological UV 
background \citep{haardtmadau96} using the {\tt CLOUDY} code 
\citep[ver. 96b4;][]{ferlandetal98}.  In constructing these 
rates the collisional and UV ionization equilibria are explicitly
calculated. This allows the direct determination of the fraction of 
ionized electrons in each mesh cell, $\chi$, rather than assuming a 
universally averaged value. 

One of the main effects of cooling and star formation is to lower the
volume-averaged gas density significantly below the cosmic baryon
density due to the conversion of gas into stars. At $z = 4$ the gas fraction 
is $\rho_{\rm g} / \rho_{\rm bar} = 0.99$ (where $\rho_{\rm bar}$ is the
mean baryon density of the Universe), whereas by $z = 0$ it has fallen
to 0.8.  As noted by \citet{rudd08}, the conversion of gas into stars
in the simulation is significantly more efficient than is observed --
especially in the highest density regions -- due to the well-known
{\em over-cooling problem} \citep[see also][]{kravtsov05,
  kravtsov09}. At the high-mass end, the halos in this simulation have
a stellar mass fraction at $z = 0$ that is approximately 50\% greater
than observed for group and cluster-mass objects \citep{gonzalez07,
  giodini09}. We note that, while the star-formation rate at high
redshift $z>4$ is likely to be {\em underestimated} \citep[as many
  halos are poorly resolved at these epochs,][]{springel03}, Figure
\ref{fig:ksz_contours} indicates that the majority of the kSZ power at
$\ell = 3000$ is sourced at lower redshift. Overall, the L60CSF
simulations should therefore underestimate the amplitude of the gas
density -- and thus the kinetic SZ -- power spectrum. On the
other hand, the absence of these processes in our non-radiative
simulations results in a mean gas density in halos that is
substantially larger than that observed. The BolshoiNR simulation
should thus over-estimate the kinetic SZ power. Therefore, while
neither simulation represents the real Universe, we expect them to
provide well-motivated lower and upper limits to theoretical estimates
of the kSZ power spectrum. These limits can be utilized to place
constraints on the kSZ signal from reionization.

To compare with these two limiting cases we also reran the L60CSF
simulation having turned off radiative cooling at $z = 1.8$
(henceforth referred to as L60CSFz2). While this run should be
considered unrealistic, it serves two useful purposes. Firstly, it
allows us to investigate whether a radical change in the
star-formation prescription can significantly modify the shape of the
window function, $W^2(k)$, and the kSZ power spectrum away from that
predicted by either the BolshoiNR or L60CSF simulations. This then
enables us to evaluate how robust our estimate of $W^2(k)$ is to
changes in baryon physics: does it vary smoothly between the
non-radiative and over-cooling case, or does it depend sensitively on
the star-formation prescription? Secondly, the final stellar (and gas)
mass fraction in halos of mass $M_{\rm 500} > 10^{12} \hmsun$ at $z =
0$ in this simulation lies between the values produced by the
BolshoiNR and L60CSF simulations. For example, for halos of mass
$M_{\rm 500} = 5\times 10^{13} \msun$, the ratio of stellar (gas) mass
to total mass in the L60CSFz2 simulation is approximately 0.12 (0.8)
of the universal baryon fraction, compared to 0 (0.9) for BolshoiNR
and 0.6 (0.45) for L60CSF. Therefore, this run provides a useful
intermediary case between the full cooling plus star-formation and
non-radiative cases. In Section \ref{sec:discussion} we discuss future
work that is required to make precision estimates of the kinetic SZ
power spectrum.

The BolshoiNR and L60CSF are the primary simulations used to gauge the
effects of cooling and star formation and to calibrate our models. As
indicated in Table~\ref{table:simulations}, the BolshoiNR simulation
has the largest box size (250 Mpc/$h$). The range of scales resolved
by this simulation encompasses the wavenumbers $0.03 \leq k \leq 100
\hmpc$ (we conservatively use an upper limit of $1/8$ the Nyquist
wavenumber corresponding to the spatial resolution). Mildly overdense
structures ($\delta \approx 10$) are followed at a grid resolution of
approximately $60 \hkpc$, corresponding to an upper limit on $k$ of about
7 $\hmpc$. Therefore, this simulation has adequate resolution in low
density regions to account for the potential contribution of
filamentary structures to the kSZ signal
\citep{atrio-barandela08}. The L60CSF simulation has a box size of
$60$ Mpc/h, resolving fluctuations over the range $0.1 \leq k \leq 110
\hmpc$. The resolution in lower-density regions is approximately $200
\hkpc$, so it is possible that density fluctuations in these regions
are slightly underestimated. However, \citet{hallman09} demonstrate
that the contribution of regions of density $\delta < 50$ to the kSZ
power spectrum in their simulations is an order of magnitude below
that of denser regions.  Therefore, comparing with the upper panel of
Figure \ref{fig:ksz_contours}, it is clear that our simulations
adequately resolve density fluctuations over the range of scales that
contribute the bulk of the kSZ signal.

While we do not directly present results from the L60NR simulation, we
use this to check the effect of varying the simulation box size (by
comparing BolshoiNR and L60NR) and separate it from the effect of
adding baryonic physics (by comparing L60NR and L60CSF).

\section{Results}
\label{sec:results}

\subsection{Gas Window Function}

\begin{figure*}
%\plotone{window_function_bolshoi.eps}
\includegraphics[scale = 0.26]{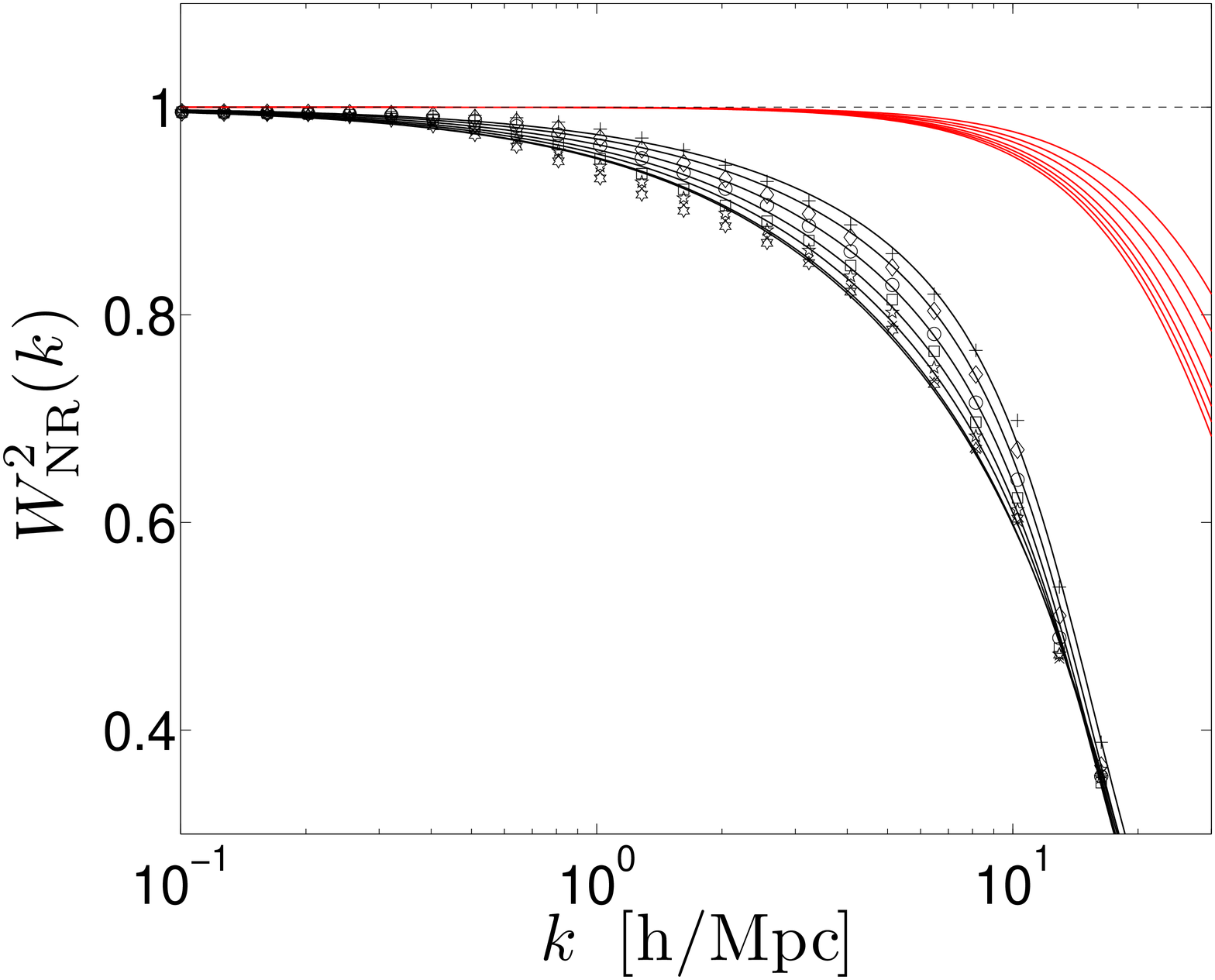}
\includegraphics[scale = 0.26]{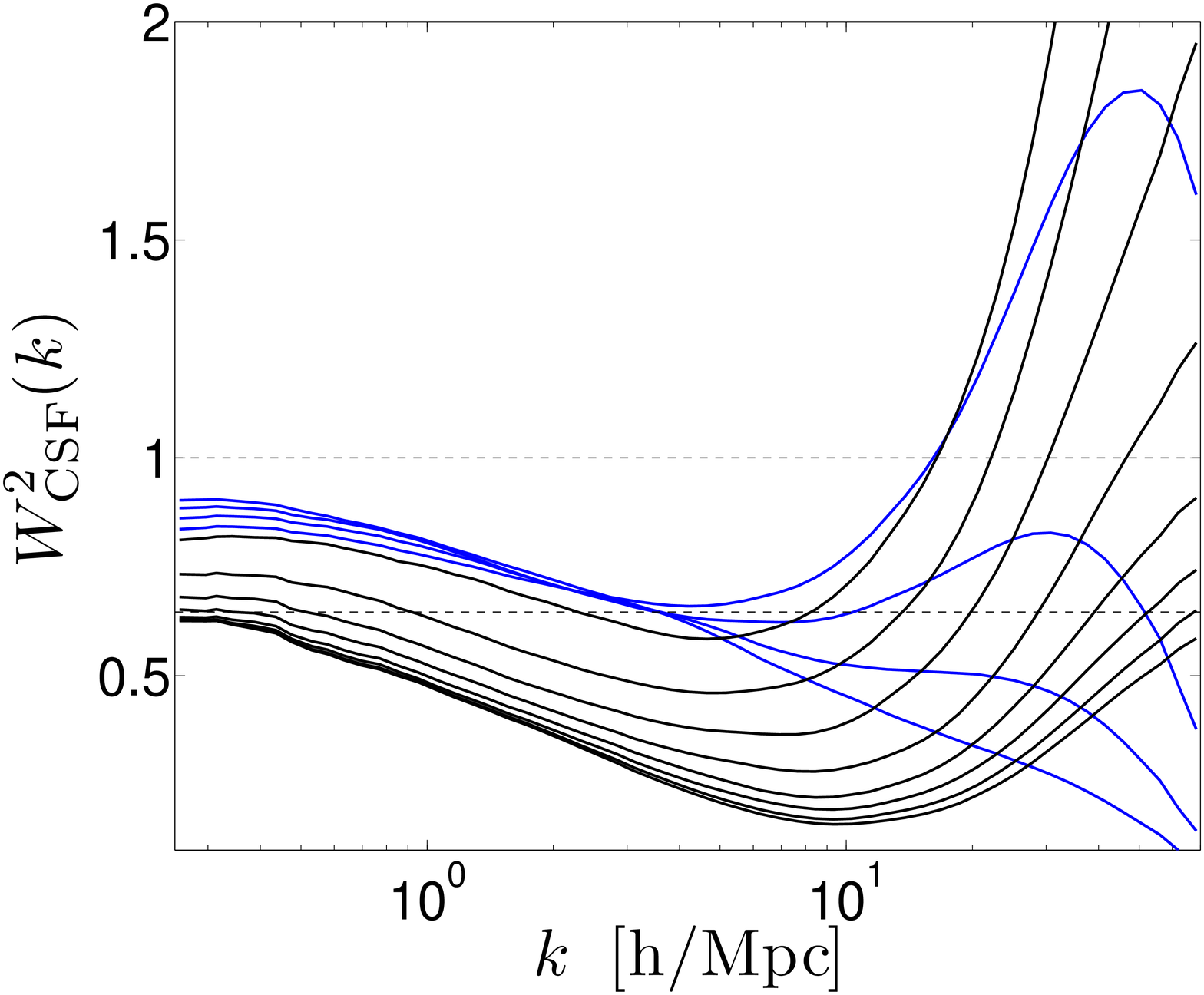}
\caption{({\bf left}) The ratio of the gas density to dark matter
  density power spectra, $W^2(k)$, measured in the BolshoiNR
  simulation (black symbols) and the linear perturbation theory
  prediction of \citet{gnedin98} (red solid lines). For each, the
  results are shown from $a = 0.4$ to $1$ (where $a$ is the scale
  factor for the size of the universe) in steps of $\Delta a = 0.1$
  (from top to bottom at $k=2\hmpc$). The black solid lines show the
  fit to the simulation points given by Equation \ref{eq:wk_nr}. ({\bf
    right}) The ratio of the free electron density to dark matter
  density power spectra in the L60CSF simulation (black lines). The
  results are shown from $a = 0.3$ to $1$ in steps of $\Delta a = 0.1$
  (from top to bottom). Also plotted is the ratio of the free electron
  density to dark matter density power spectrum for the L60CSFz2
  simulation (in which radiative cooling is turned off at $z = 1.8$) at
  $a = 0.4, 0.5, 0.68$ and $1$ (from top to bottom at $k = 10\hmpc$)}
\label{fig:wk_bolshoi}
\end{figure*}

In the left panel of Figure \ref{fig:wk_bolshoi} we plot the window
function $W^2(k) = P_{\rm gas}(k)/P_{\rm DM}(k)$ for the BolshoiNR
simulation, where $P_{\rm gas}$ and $P_{\rm DM}$ are the measured gas
and dark matter density power spectra. The points show the results
over a range of timesteps, corresponding to $a = 0.4$ to $1.0$ in
steps of 0.1 (from upper to lower at $k=2\hmpc$).  The black solid
lines are fits to the simulation results using the fitting function
given below. The red lines show the linear-theory window function of
\citet{gnedin98} given by Equation \ref{eq:wk_lin}.

Fluctuations in the gas density follow those of the dark matter at
large scales ($k < 1 \hmpc$), but are rapidly suppressed towards
smaller scales as the gas thermal pressure begins to counter gravity
in overdense regions.  Note that the largest value of $k$ plotted is
significantly below the maximum that is resolvable for the simulation. Hence,
the truncation of power at small scales in $P_{\rm gas}$ is due to a
physical smoothing of the gas density by thermal interactions rather
than an artificial smoothing due to the finite resolution of the
simulation.

We find that the following fitting-function provides a good description of the BolshoiNR simulation results, 
\beq 
W_{\rm NR}(k,a) = 0.5\left(\exp{(-k/k_f')} + \frac{1}{1 + (g(a) k / k_f')^{7/2}}\right)
\;,
\label{eq:wk_nr}
\eeq 
where $k_f' = 12.6 a^{-1} + 6.3$ and $g(a) = 0.84 a^{-1}$. We
have verified that Equation \ref{eq:wk_nr} also provides a good match
to the window function measured from the L60NR simulation. The
difference between the BolshoiNR and L60NR simulation box sizes and
cosmological parameters do not significantly affect our results.

At the final output, the characteristic filter scale, $k_f'$, is
approximately a factor of $3.3$ less than that of the linear-theory
prediction given by Equation \ref{eq:kf}. This is also evident from
comparing the window function suggested by \citet[][red
  lines]{gnedin98} with the simulation results; the smoothing of the
gas density fluctuations occurs at much smaller scales than in the
simulation. The difference between the \citet{gnedin98} window
function and the simulation results is due to shock heating of the gas
in the simulation as density perturbations become non-linear. To
calculate the Jean's scale in Equation \ref{eq:kf} we assume that the
gas is initially coupled to the CMB temperature, but then evolves
adiabatically, $T_{\rm gas} \propto 1/a^2$, until it reaches a minimum
temperature of 300K (which is also imposed in the simulation). The
lower gas temperature at late times results in a large Jeans
wavenumber, and thus the smoothing of the gas density is limited to
very small scales (e.g., $ \{k_J, k_f\} \approx \{32,63\} \hmpc$ at $a
= 1$). However, in the simulation, as halos collapse and grow,
accretion shocks heat much of the gas to temperatures significantly
above the 300K temperature floor (note that this simulation, unlike
the L60CSF run, does not include a UV background). The mean
(mass-weighted) gas temperature in the simulation is in excess of
$10^6$ K at $z = 0$. Thus we see smoothing of the gas distribution at
much larger scales in the simulation than predicted by purely
adiabatic evolution.

On the right side of Figure \ref{fig:wk_bolshoi} we show the window
function, $W_{\rm CSF}(k)$, measured in our L60CSF simulation at the
timesteps corresponding to $a = 0.3$ to $1.0$ in steps of $0.1$ (black
lines, from top to bottom at $k = 1 \hmpc$; for clarity we represent
the simulation results with lines rather than symbols). In this
simulation we are able to measure the ionization fraction of hydrogen
in each cell. We therefore plot the ratio of the free-electron (rather
than the gas) density power spectrum to that of the dark
matter. However, since the global neutral fraction is very small,
using the gas density produces similar results.

The inclusion of gas cooling and star-formation produces a very
different window function to the non-radiative simulation.  Firstly,
the overall amplitude at large scales is clearly time-dependent. At $k
= 0.2 \hmpc$, $W^2(k)$ varies from 0.92 at $a = 0.2$ to 0.62 at $a =
1.0$. For the BolshoiNR simulation we found that $W_{\rm NR}^2(k=0.2)
\approx 1$ at every timestep. The difference is due to the evolving
gas fraction within halos in the CSF simulation. At $a = 1$, the
square of the ratio of the gas to total baryon density in the box is
$(\rho_{\rm g} / \rho_{\rm bar})^2 = 0.64$ (shown as the horizontal
dot-dashed line).  The low gas density within halos in this simulation
significantly reduces the power spectrum of gas density fluctuations
relative to the dark matter. This is particularly evident at low $k$
as the two-halo contribution to the gas density power spectrum is
proportional to the square of the mean gas density in halos
\citep{semboloni11}. Therefore, as the gas density is reduced, so is
the amplitude of $W_{\rm CSF}^2(k)$ at large scales.

At intermediate scales $W^2(k)$ slowly decreases, but starts rising
again at $k > 5 \hmpc$. The increase in power towards small scales is
due to the cooling of gas at the center of halos, steepening the
density profile at small radii relative to the dark matter and thus
boosting the small-scale power. For earlier outputs (upper lines) we
see that the power in the free-electron density can greatly exceed
that of the dark matter ($W_{\rm CSF}^2(k) > 1$). With time, this bias
gradually disappears due to two effects. Firstly, as the simulation
progresses, the cold gas in halo cores is converted to
stars. Secondly, the gas in the very center of halos can cool to the
point at which hydrogen is no longer ionized, reducing the
free-electron density. However, the neutral fraction always remains
small and thus the former is the dominant effect. Using the gas
(rather than free-electron) power spectrum produces very similar
results.

The blue lines in the right-hand panel of Figure \ref{fig:wk_bolshoi}
show the window function obtained for the L60CSFz2 simulation, in
which radiative cooling is artificially turned off at $z_{\rm off} =
1.8$. The results are given for the timesteps corresponding to $a =
0.4, 0.5, 0.68 $ and 1 (from top to bottom at $k = 10
\hmpc$). Switching off radiative cooling has a clear effect on the gas
distribution across the range of scales probed. After $z_{\rm off}$,
the cold, dense clumps of gas in halo cores begin to heat and expand,
rapidly suppressing small-scale density fluctuations. The reduction of
small-scale power thus occurs in both CSF simulations, but for very
different reasons: in L60CSF cold gas is converted into stars, whereas
in L60CSFz2 the gas mixes with its warmer surroundings and begins to
expand. This also results in a different behavior at large spatial
scales. In L60CSFz2, the gas becomes less centrally concentrated in
halos, reducing the suppression in power (relative to the dark matter)
that is seen in the L60CSF simulation. As the simulation progresses
towards $z = 0$, $W^2(k)$ thus increases slightly at small $k$. By the
final output, the shape of the window function is beginning to
resemble that of the BolshoiNR simulation (left-hand panel of Figure
\ref{fig:wk_bolshoi}). Therefore, once cooling is turned off, the window
function obtained from the L60CSFz2 simulation begins to transition
from a CSF-like shape to a NR-like shape. We show the effect of this
on the kSZ power spectrum in Section \ref{sec:ksz}.

The form of $W_{\rm CSF}^2(k)$ is evidently more complicated than
$W_{\rm NR}^2(k)$. To incorporate the CSF window functions into our
calculation of the kinetic SZ power spectrum, we instead interpolate
between the simulation results for the redshifts between outputs.  We
assume that $W^2(k)$ remains fixed at its value at $k = 0.2 \hmpc$ for
scales with $k< 0.2 \hmpc$.  For epochs earlier than $a = 0.3$ ($z >
2.33$), we assume that $W_{\rm CSF}^2(k)$ smoothly converges towards
unity at all scales.

\subsection{Momentum Power Spectrum}
\label{sec:momentum}

\begin{figure*}
\plotone{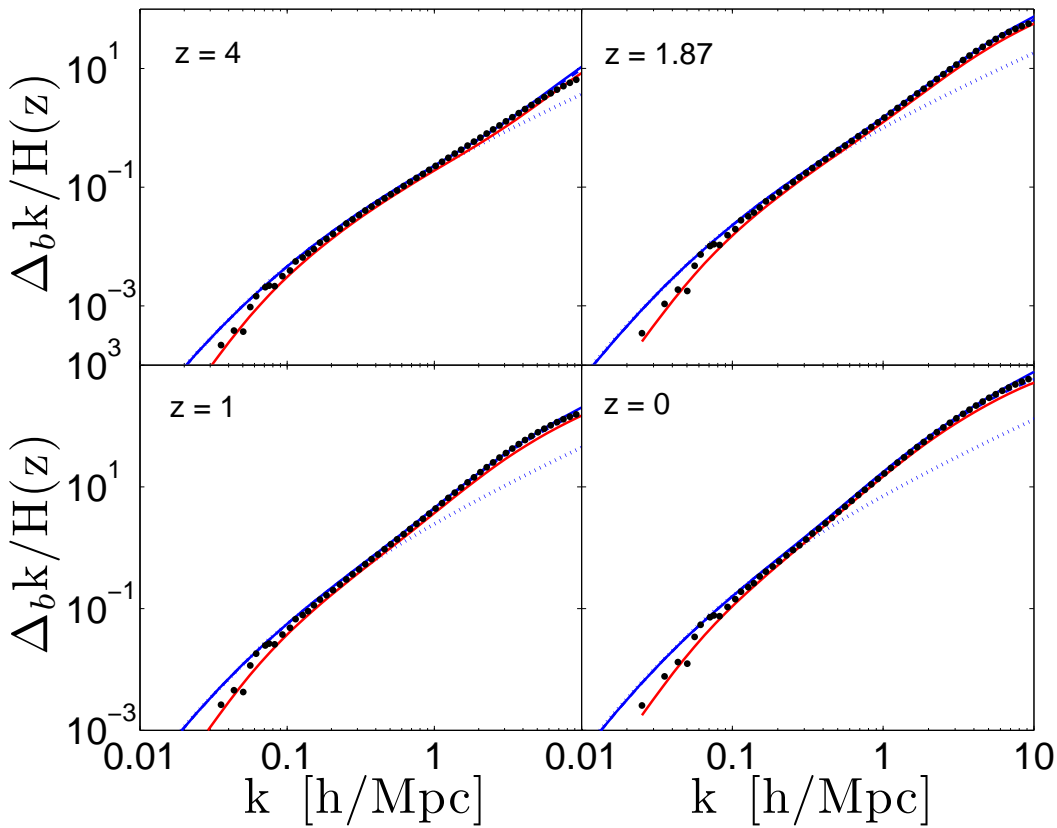}
\caption{The power spectrum of the curl component of the momentum
  field in the BolshoiNR simulation at four timesteps: $z = 4$, 1.87,
  1 and 0 (top left, top right, lower left, lower right). The black
  dots represent the simulation results, the blue lines give the
  analytic prediction of $\Delta_B k / H(z)$ in two regimes: the linear
  regime calculation given by Equation \ref{eq:deltab_lin} (dotted),
  and the non-linear version given by Equation \ref{eq:deltab_seminl}
  (solid). The red line shows the `truncated' model (see text).}
\label{fig:deltab_nr}
\end{figure*}

\begin{figure*}
\plotone{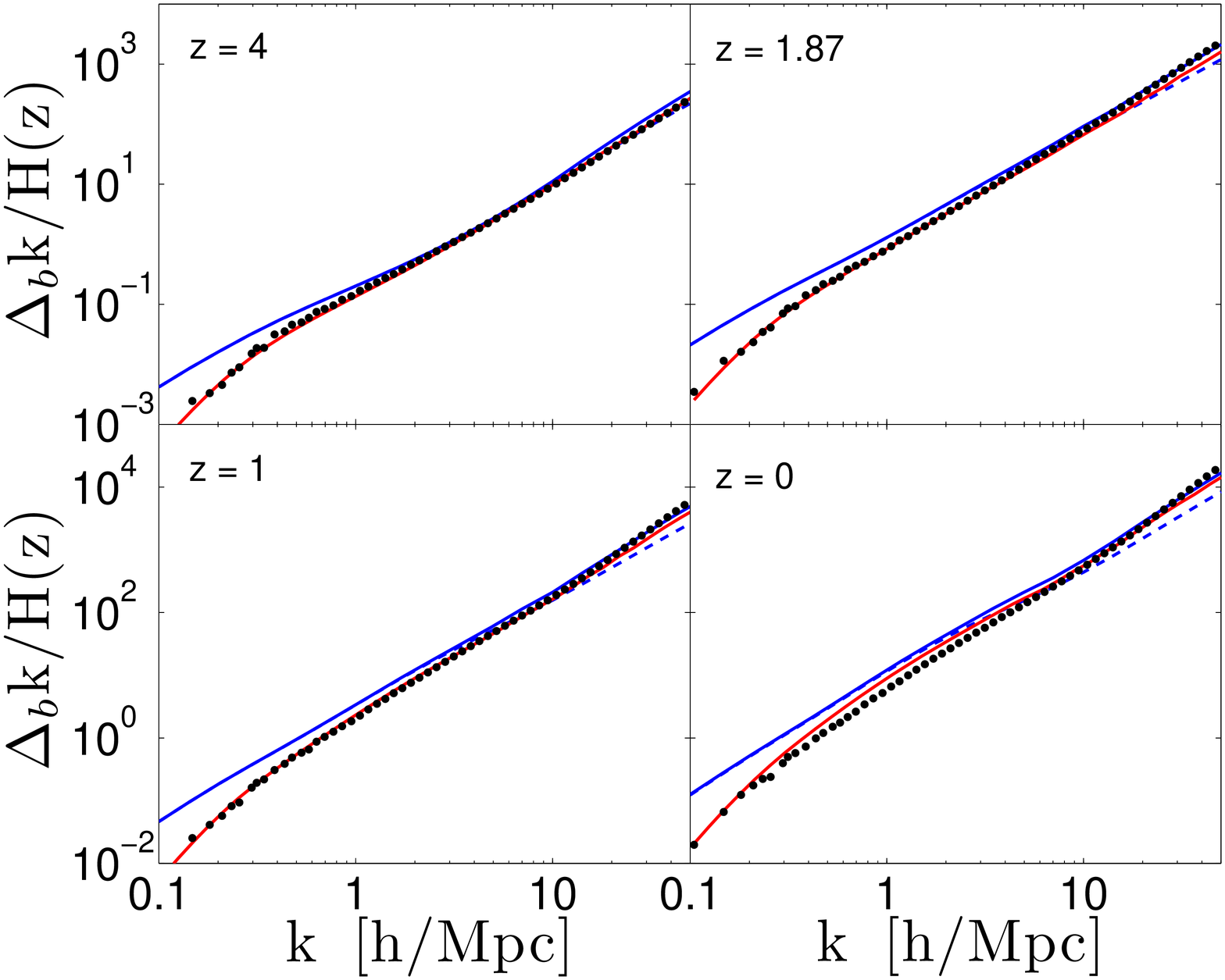}
\caption{Similar to Figure \ref{fig:deltab_nr} but for the L60CSF
  simulation. The black dots represent the simulation results, the
  solid and dashed blue lines give the analytic prediction of
  $\Delta_B k / H(z)$ including and omitting the halo contraction
  correction, respectively. The red line shows the `truncated' model
  (see text).}
\label{fig:deltab_csf}
\end{figure*}

We now compare the power spectrum of the curl component of the gas
momentum field, $\Delta_B^2$, predicted by our model with that
measured directly from the simulations. This comparison enables us to
test whether our model accurately captures the impact of non-linear
structure growth and baryonic physics in the simulations. This is of
particularly importance as the kSZ power spectrum is a weighted
integral of $\Delta_B^2$ over redshift. We also use our model to
demonstrate the impact on $\Delta_B^2$ of the truncation of
large-scale velocity modes due to the finite size of our simulated
volumes. 

To measure $\Delta_B^2$ in our simulations we use a $1024^3$ mesh
rather than the fully refined mesh in each simulation. Hence, we are
limited to a narrower range of scales than for the density power
spectra. For the BolshoiNR simulation (which has a box side-length of
250 Mpc/$h$) this range is $0.03 \leq k \leq 12 \hmpc$. For the L60CSF
simulation (60 Mpc/$h$ box) the range is $0.1 \leq k \leq 50$.

In Figure \ref{fig:deltab_nr} we plot $\Delta_B^2$ for the BolshoiNR
simulation at four timesteps, corresponding to $z = 4$, 1.87, 1 and
0. The momentum power is shown in terms of the dimensionless quantity
$\Delta_B k / H(z)$. The black dots represent the measured simulation
power spectrum.  The blue lines give the analytic prediction in two
regimes: the linear regime calculation (the Ostriker-Vishniac
effect) given by Equation \ref{eq:deltab_lin} (dotted), and our
fiducial model given in Equation \ref{eq:deltab_ksz}, which includes
the non-linear corrections to the density power spectrum (solid).
Both include the window function, $W_{\rm NR}^2(k)$.

Non-linear density fluctuations become significant when $\Delta_b k /
H(z) \approx 1$. Our full non-linear model reproduces the simulation
results extremely well at $k > 0.1 \hmpc$, but lies systematically
above the simulation points at lower $k$. This is simply due to the
finite volume of the simulation box. $\Delta_B^2(k)$ is generated by a
convolution between velocity modes at wavenumber $k'$ with density modes
at $|k-k'|$. The largest contribution comes from velocity modes at
large scales coupling to density modes at smaller scales. Velocity
modes on scales larger than the simulation box size are not accounted
for, resulting in an underestimate of the momentum power.

To demonstrate that this is indeed the case, we recalculate
$\Delta_B^2$, placing an upper and lower limit on the integral over
$k'$ in Equation \ref{eq:deltab_ksz} such that only the velocity modes
encompassed by the simulation box are included. The red lines in each
panel of Figure \ref{fig:deltab_nr} show the results of this
`truncated' model. It accurately reproduces the simulation on all
scales. This was previously demonstrated by \citet{zhang04}, who
performed a similar test by calculating a discretized version of
Equation \ref{eq:deltab_ksz} and compared the results with their
simulations. Despite the overall very good agreement, the truncated
model does slightly underestimate the simulation results at $k > 2
\hmpc$. This may be due to the impact of a non-linear, curl component,
of the velocity field generated by shell-crossing during halo
formation \citep[as suggested by][]{zhang04, shao11}. To mitigate
this, \citet{zhang04} suggest replacing $\Pdd$ in Equation
\ref{eq:deltab_seminl} with it's non-linear counterpart. When we do
so, we find that the ratio of model to simulation increases slightly,
from 0.88 to 0.94 at $k = 2 \hmpc$.

In Figure \ref{fig:deltab_csf} we plot the curl component of the gas
momentum power spectrum in the L60CSF simulation (black dots) at the
same four timesteps.  In this figure, the blue lines show the results
omitting (dashed) and including (solid) the halo contraction
correction applied to the (non-linear) density power spectrum
described in Section~\ref{sec:nonlinear_corrections}.  Both lines
include the window function $W_{\rm CSF}^2(k)$ measured in the L60CSF
simulation.

At the low-$k$ end, the simulation points again lie systematically
below the model. However, for this simulation the discrepancy extends
to much higher $k$.  The L60CSF simulation box is more than a factor
of four times smaller than the BolshoiNR simulation (see Table
\ref{table:simulations}), and so the velocity mode truncation is more
severe. We note that our L60NR simulation underestimates $\Delta_b^2$
(relative to our non-radiative model) by a similar amount, confirming
that the box-size effect is independent of the baryonic physics
included in the simulation. As in Figure \ref{fig:deltab_nr}, the red
lines show our model prediction having limited the calculation to
include only those scales encompassed by the simulation. The model
again reproduces the simulation results very well (although at $a = 1$
it appears to slightly overestimate the momentum power at scales
around $k = 1 \hmpc$).

The dashed lines demonstrate the model with no halo contraction
correction. It is clear that, without this correction, the model
underestimates $\Delta_B^2$ for wavenumbers $k > 10 \hmpc$, especially
at $z = 0$. The solid blue line demonstrates that our simple
correction does a reasonable job of reproducing the simulated momentum
power spectrum. However, as Figure \ref{fig:ksz_contours}
demonstrates, these scales do not contribute significantly to the kSZ
power spectrum at $\ell = 3000$.

\section{Kinetic SZ Power Spectrum}
\label{sec:ksz}

We now utilize the gas window functions measured in our NR and CSF
simulations to explore the impact of baryonic physics on the shape and
amplitude of the kinetic SZ power spectrum. As described in Section
\ref{sec:final_model}, we evaluate three baseline models: DM, NR and
CSF. For the DM matter model $W^2(k) = 1$, for the NR and CSF we
use the window functions $W_{\rm NR,CSF}^2(k)$ measured in our
simulations. We also investigate the redshift dependence of the kSZ
power spectrum, and its scaling with cosmological parameters for each
model. Finally we compare the results presented in this work with
those of previous theoretical studies as well as the latest
observational constraints.

\subsection{The Impact of Baryon Physics}

\begin{figure}
%\plotone{f5.eps}
\includegraphics[scale = 0.26]{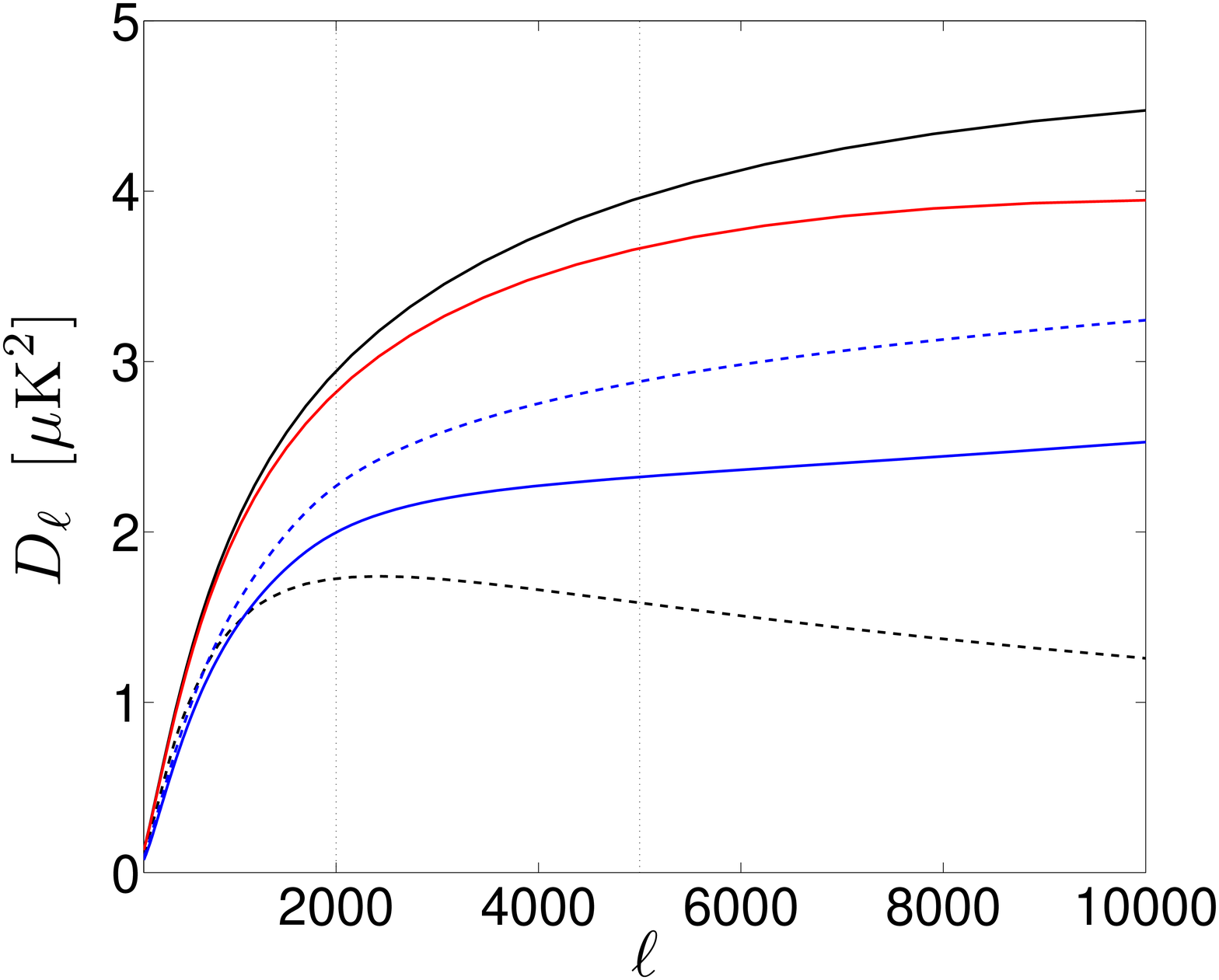}
\caption{The kinetic SZ power spectrum for our three models: DM, NR
  and CSF (solid black, red and blue lines). The black dashed line
  shows the linear theory version of the DM model (the
  Ostriker-Vishniac effect). The blue dashed line shows kSZ power
  spectrum calculated using the L60CSFz2 window function. The vertical
  dotted lines denote the approximate range in $\ell$ over which
  current small-scale CMB experiments such as SPT and ACT are
  sensitive to kSZ power.}
\label{fig:ksz}
\end{figure}

In Figure \ref{fig:ksz} we plot the kinetic SZ power spectrum for our
three models: DM, NR and CSF (solid black, red and blue lines).  The
dashed lines represent variants to the DM and CSF models. The black
dashed line shows the linear theory version of the DM model, which we
label OV (Ostriker-Vishniac). The blue dashed line shows the kSZ power
spectrum calculated using the L60CSFz2 simulation window function. The
power spectra are plotted in terms of $D_{\ell} = \ell (\ell+1)
C_{\ell} / 2\pi$.

Comparing the dashed and solid black lines demonstrates the impact of
non-linear structure growth on the kinetic SZ power spectrum.
Including the non-linear corrections significantly boosts kSZ power at
all but the largest angular scales. At $\ell = 3000$, non-linear
corrections have increased the kSZ signal by a factor of 2. By $\ell =
10,000$, this has increased to a factor of $3.5$. The kSZ effect is
sourced by small-scale density fluctuations caught up in large-scale
bulk velocity flows. Therefore, the large boost in the amplitude of
small-scale density perturbations due to non-linear gravitational
collapse also enhances the kSZ power, especially at small angular
scales.

The difference between the DM and NR models are small. At $\ell =
3000$ $(10,000)$ the BolshoiNR window function reduces the kSZ power
from the DM model by $0.20$ $(0.72) \mksq$. The NR window function
suppresses power at scales $k \gtrsim 8 \hmpc$ (see the left panel of Figure
\ref{fig:wk_bolshoi}). As demonstrated by
Figure~\ref{fig:ksz_contours}, gas density fluctuations at these
scales contribute little to the kSZ power spectrum (at $\ell =
3000$). The application of $W_{\rm NR}^2(k)$ thus translates into only a
small reduction of power (relative to the DM model) at the angular
scales most sensitively probed by current small-scale CMB experiments.

The window function measured in the L60CSF simulation has a greater
impact on the kSZ power spectrum. The large-scale suppression of the
gas density power spectrum seen in
Figure~\ref{fig:wk_bolshoi} produces a significant reduction in kSZ
power at all angular scales. At $\ell = 3000$, the power is reduced by
$1.05 \mksq$ (32\%) and by $1.43 \mksq$ (36\%) at $\ell = 10,000$,
relative to the NR model. As discussed in the previous section, this
reduction in power is primarily driven by the decreased gas density in
halos due to over-cooling and excessive star-formation. A similar
reduction in kSZ power was noted by \citet{trac11} when the stellar
mass fraction in groups and clusters was increased in their
semi-analytic model, reducing the gas fraction by an equivalent
amount. 

The blue dashed shows the results using the window function measured
in the L60CSFz2 model. The amplitude of the power spectrum in this
case is close to the mean of the NR and CSF models, with the shape
unchanged. A comparison to Figures \ref{fig:ksz_contours} and
\ref{fig:wk_bolshoi} provides a clear explanation. Figure
\ref{fig:ksz_contours} demonstrates that the peak of the kSZ
contribution comes from spatial scales corresponding to $1 \hmpc < k <
3 \hmpc$ at redshift $z \sim 1$. At these scales and redshifts, the
L60CSFz2 window function lies almost directly between that of the
BolshoiNR and L60CSF simulations, and thus the kSZ power spectrum also
lies in between these two cases. The dominant factor determining
the amplitude of kSZ power thus appears to be the mean gas density in
halos, which is controlled by the star-formation rate.

We have also investigated the impact of non-linear velocity
fluctuations by replacing the linear density power spectrum (i.e. the
velocity component of the $\Delta_B^2$) with its non-linear
counterpart, as suggested by \citet{zhang04}. We find that, at $\ell =
3000 \; (10,000)$, the amplitude of the kSZ power spectrum in the NR
model increases by 1\% (4\%). We have chosen to omit this correction
from our fiducial models, as the velocity contribution to $\Delta_B^2$
in Equation \ref{eq:deltab_seminl} is determined by the gradient of
the density field and is therefore curl-free. Hence, it is not clear
how one should correctly account for non-linearities in the velocity
field. Nevertheless, the impact on the kSZ power spectrum is likely to
be small.

\begin{figure}
%\plotone{f6.eps}
\includegraphics[scale = 0.27]{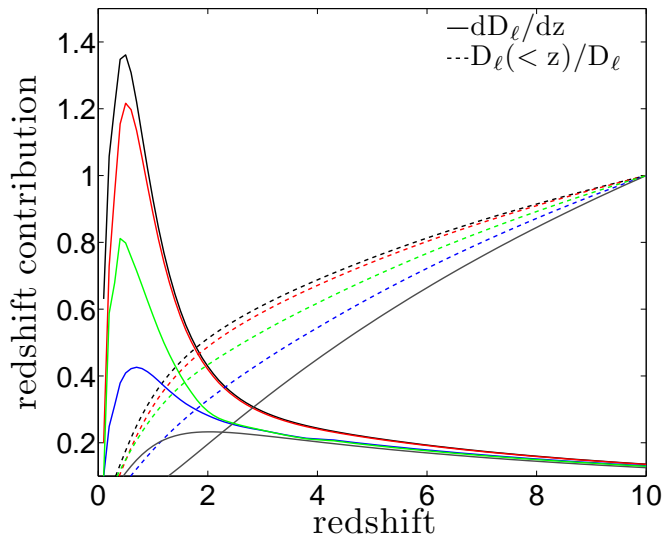}
\caption{The contribution of slices in redshift to the kSZ power
  spectrum at $\ell = 3000$. The solid lines show the differential
  contribution, the dashed lines show the cumulative contribution. The
  line colours represent our different models, DM (black), NR (red),
  CSF (blue), and the OV contribution to the DM model (grey). Also
  plotted is the kSZ redshift distribution predicted by the L60CSFz2
  simulation (green).}
\label{fig:ksz_redshift}
\end{figure}

In Figure \ref{fig:ksz_redshift} we plot the relative contribution to
the kSZ power spectrum of slices in redshift over the range $0 \leq z
\leq z_{rei}$, at $\ell = 3000$. The solid lines show the differential
contribution, $\drm D_\ell / \drm z$, the dashed lines show the
cumulative contribution, $D_\ell(<z) / D_\ell(<z_{rei})$, where
$z_{rei} = 10$. The line colors represent our different models, DM
(black), NR (red), CSF (blue) and also the OV contribution to the DM
model (grey). For comparison, we also plot the results obtained using
the L60CSFz2 window function (green).

Comparing the differential redshift contribution of the DM and OV
models again demonstrates the enhancement provided by non-linear
structure formation at low redshift. Half of the kSZ signal in the DM
model comes from $z < 2$, whereas the equivalent fraction for the OV
is attained at much higher redshifts ($z \sim 5$).

The NR model has a similar redshift distribution to the DM model.
However, the CSF model predicts a redshift distribution that is
significantly flatter. While the differential redshift distribution
peaks at the same redshift, the amplitude of the peak is considerably
lower. The half-way point of the cumulative distribution is at a much
higher redshift, $z \sim 4$, than for the DM and NR models. The right
panel of Figure~\ref{fig:wk_bolshoi} shows that the effects of cooling
and star-formation, which reduce the gas  density
in halos, become increasingly more significant towards lower
redshift. This counteracts the boost to the kSZ signal provided by
non-linear density fluctuations such that the redshift distribution of
the CSF more closely resembles that of the OV model. 

The L60CSFz2 simulation, in which radiative cooling is turned-off at
$z = 1.8$, produces a kSZ redshift distribution that lies almost
exactly in between the NR and CSF models. The distribution diverges
from the L60CSF line at $z = 1.8$ and then remains at roughly 2/3 the
amplitude of the NR model. This is to be expected; once cooling is
turned off, gas is no longer being consumed by star-formation. Hence,
at lower redshifts, the differences between the L60CSFz2 and the NR
model cease to grow. These results suggest that other (and less
sudden) variations in the star-formation history would produce a
predictable change in the kSZ redshift distribution: the decrease
compared to the NR case would depend on the integrated star-formation
rate to that redshift. While it may underestimate the star-formation
rate at high redshift (due to the limited mass resolution), the CSF
model only diverges significantly from the OV prediction at $z <
4$. Given the over-cooling problem at low redshift, it is reasonable
to assume that the NR and CSF cases encompass the expected redshift
distribution of the kSZ power spectrum at $\ell = 3000$.

The redshift distribution of the kSZ power spectrum can potentially be
measured using kSZ tomography \citep{ho09, shao11}. The tomography
method utilizes catalogues of galaxies with precisely measured
spectroscopic redshifts to reconstruct the large-scale velocity field
from the 3-D galaxy distribution. When integrated along the line of
sight, the product of the reconstructed velocity and density fields
provides an estimator for the kinetic SZ temperature
fluctuations. This estimator can be broken down into redshift slices
and cross-correlated with a map at CMB frequencies enabling the true
kSZ contribution from within that redshift slice to be measured.

\citet{shao11} demonstrate that, by combining Planck with BigBOSS
\citep{schlegel11} it will be possible to measure the kSZ power from
the redshift range $0.2 \leq z \leq 0.6$ to better than 10\% precision
at $\ell = 3000$, and within the range $0.6 \leq z \leq 1$ to better
than 20\%. Utilizing data from higher angular resolution CMB
experiments such as ACT should significantly increase the precision of
these measurements. This will not only provide a unique,
temperature-independent probe of the inter-galactic medium at lower
densities than those typically probed by X-ray and tSZ surveys,
but would place tight constraints on the models of the kSZ power
spectrum presented here. Figure \ref{fig:ksz_redshift} demonstrates
that much of the theoretical uncertainty on the kSZ signal comes from
within the range $0 \leq z \leq 2$; kSZ tomography can provide a
powerful probe of the kinetic SZ effect over the lower half of this
redshift range.

\subsection{Cosmological scaling}
\label{sec:cosm_scale}

We have discussed in detail the theoretical uncertainty in the kinetic
SZ power spectrum from astrophysical sources. However, the kSZ signal
also scales sensitively with cosmological parameters. There is
therefore an additional uncertainty on theoretical estimates of the
kSZ power spectrum due to the precision with which these parameters
have been measured. In this section we investigate how the amplitude
of the kSZ power spectrum scales with cosmological parameters -- in
particular, $\sigma_8$ and $\tau$ -- and how this scaling varies
between our NR and CSF models.

We assume that $D_\ell$ scales as a power-law with each cosmological
parameter $p$, i.e., $D_{\ell} \propto p^\alpha$. We vary each
parameter by 20\% of its fiducial value (holding the other parameters
fixed) and measure the value of $\alpha$. We have verified that the
power-law scaling is appropriate for all parameters. The results are
presented in Tables \ref{tab:cosmscaling_NR} and
\ref{tab:cosmscaling_CSF}. We give the results over the angular
multipole number range $1000 < \ell < 10,000$. Also given are the
band-powers, $D_\ell$, for each model.

The kSZ power spectrum amplitude is most sensitive to $\sigma_8$,
scaling as $\sigma_8^{4.3 - 5.1}$ from $\ell = 1000$ to $10,000$ (NR
model). This scaling is simple to understand. The kSZ power spectrum
involves the product of the density and velocity power spectra; in the
linear regime each contributes a power of 2 to the scaling with
$\sigma_8$.  The OV power spectrum thus scales exactly as
$\sigma_8^4$. Non-linear corrections to the matter power spectrum
steepen the $\sigma_8$ dependence. This steepening is related to the
fraction of the kSZ signal that originates at low redshift where the
non-linear corrections have grown large. Figure \ref{fig:ksz_redshift}
shows that the maximal contribution to the OV signal comes from $z =
2$, whereas the full kSZ signal for the DM model peaks at $z =
0.5$. This shift in the redshift contribution is driven by non-linear
structure formation, which explains the steeper $\sigma_8$
dependence. The same argument applies to the difference in the
$\sigma_8$ scaling between the NR and CSF models; a smaller fraction
of the signal derives from low redshift in the latter and so the
scaling with $\sigma_8$ is weaker. We note that changing $\sigma_8$
principally varies the overall amplitude of the kSZ signal; while the
scaling does have some $\ell$-dependence, the impact on the shape of
the power spectrum is small over the parameter space explored.

Some of the tightest constraints on $\sigma_8$ to date have been
obtained from measurements of the primary CMB power spectrum and the
abundance of galaxy clusters. For example, \citet{vikhlinin09} used a
sample of 86 clusters to obtain $\sigma_8 (\Omega_M / 0.25)^{0.47} =
0.81 \pm 0.01 \pm 0.02$ (statistical and systematic errors,
respectively). Using the CSF model scaling, this translates to an
uncertainty of approximately $\pm$15\% on the kSZ power spectrum at
$\ell = 3000$. This is equivalent to the uncertainty in $D_\ell$
provided by baryon physics presented in the previous section.

The magnitude of the kSZ signal is also sensitive to the optical depth
to reionization, $\tau$, which is directly related to the redshift at
which reionization occurs (see Equation \ref{eq:tau}). In Tables
\ref{tab:cosmscaling_NR} and \ref{tab:cosmscaling_CSF} we give the
scaling of $\dell$ with both $\tau$ and $z_{\rm rei}$. The CSF model
has a greater sensitivity to each of these parameters as a slightly
larger fraction of the signal is generated at high redshift. In
principle, the post-reionization kSZ power spectrum could be used to
measure $z_{\rm rei}$. In practice, the uncertainty on other
cosmological parameters washes out this information.

To this point we have ignored the effect of helium reionization. The
amplitude of the kSZ signal also scales as the square of the
ionization fraction, $\chi^2$, which is dependent on the ionization
state of helium. Our fiducial model assumes that helium remains
neutral at all epochs. If, instead, we assume that helium is singly or
doubly ionized at $z_{\rm rei}$ then the amplitude of the kSZ power
spectrum will increase by a ($\ell$-independent) factor of 1.16 or
1.33, due to the increase in the free electron density. A more
realistic model in which helium is singly ionized at $z = 6$ and
doubly ionized at $z = 3$ \citep[e.g.][]{furlanetto08} would increase
the kSZ power by a factor of 1.22 (1.26) at $\ell = 3000$ (10,000)
relative to our baseline, neutral helium, model. Hence, the level of
uncertainty on the kSZ power spectrum due to Helium reionization is
equivalent to that due to the uncertainty on $\sigma_8$.

\begin{table}[ht!]
\begin{center}
\caption{\label{tab:cosmscaling_NR} kSZ Cosmological Scaling: NR}
\small
\begin{tabular}{c|c|c|c|c|c|c}
\hline\hline
\rule[-2mm]{0mm}{6mm}
$\ell$ & $\dell$ & $H_0$ & $\sigma_8$ & $\Omega_b$ & $z_{\rm rei}$ & $\tau $ \\
 & $\mksq$ & 70 {\rm km/s/Mpc} & 0.82 & 0.047 & 10.0 & 0.076 \\
\hline
1000 & 1.97 & 1.18 & 4.31 & 2.29 & 0.46 & 0.32 \\ 
2000 & 2.82 & 1.51 & 4.48 & 2.18 & 0.46 & 0.32 \\ 
3000 & 3.24 & 1.68 & 4.62 & 2.13 & 0.43 & 0.30 \\ 
4000 & 3.49 & 1.80 & 4.73 & 2.10 & 0.40 & 0.28 \\ 
5000 & 3.66 & 1.88 & 4.81 & 2.08 & 0.37 & 0.26 \\ 
6000 & 3.77 & 1.95 & 4.88 & 2.06 & 0.34 & 0.24 \\ 
7000 & 3.85 & 2.00 & 4.94 & 2.05 & 0.32 & 0.22 \\ 
8000 & 3.90 & 2.05 & 4.99 & 2.03 & 0.30 & 0.21 \\ 
9000 & 3.93 & 2.09 & 5.04 & 2.02 & 0.29 & 0.20 \\ 
10000 & 3.95 & 2.13 & 5.07 & 2.01 & 0.28 & 0.19 \\
\hline
\end{tabular}
\normalsize \tablecomments{The scaling $\alpha$ of the kSZ power
  spectrum with cosmological parameters, $p$, for the NR model, where
  we assume $D_\ell \propto p^\alpha$. The fiducial parameters are
  given in the second line and the corresponding values of $\dell$ in
  the second column.}
\end{center}
\end{table}

\begin{table}[ht!]
\begin{center}
\caption{\label{tab:cosmscaling_CSF} kSZ Cosmological Scaling: CSF}
\small
\begin{tabular}{c|c|c|c|c|c|c}
\hline\hline
\rule[-2mm]{0mm}{6mm}
$\ell$ & $\dell$ & $H_0$ & $\sigma_8$ & $\Omega_b$ & $z_{\rm rei}$ & $\tau $ \\
 & $\mksq$ & 70 {\rm km/s/Mpc} & 0.82 & 0.047 & 10.0 & 0.076 \\
\hline
1000 & 1.43 & 1.09 & 4.19 & 2.31 & 0.63 & 0.43 \\ 
2000 & 2.00 & 1.46 & 4.33 & 2.18 & 0.66 & 0.45 \\ 
3000 & 2.19 & 1.65 & 4.46 & 2.12 & 0.64 & 0.44 \\ 
4000 & 2.27 & 1.78 & 4.57 & 2.09 & 0.60 & 0.41 \\ 
5000 & 2.32 & 1.87 & 4.67 & 2.06 & 0.55 & 0.38 \\ 
6000 & 2.36 & 1.94 & 4.76 & 2.04 & 0.52 & 0.35 \\ 
7000 & 2.40 & 2.00 & 4.83 & 2.02 & 0.48 & 0.33 \\ 
8000 & 2.44 & 2.06 & 4.89 & 2.01 & 0.45 & 0.31 \\ 
9000 & 2.48 & 2.10 & 4.95 & 2.00 & 0.42 & 0.29 \\ 
10000 & 2.52 & 2.14 & 4.99 & 1.99 & 0.40 & 0.27 \\ 
\hline
\end{tabular}
\normalsize \tablecomments{Similar to Table \ref{tab:cosmscaling_NR} but for the CSF model.}
\end{center}
\end{table}

\subsection{Comparisons with Simulations}

A number of previous studies have used the output of cosmological
simulations to make predictions for the shape and amplitude of the
kinetic SZ power spectrum \citep{daSilva01, white02, hallman09,
  trac11, battaglia10}.  Rather than calibrating an analytic model to
their simulations (as we have done), these studies generate mock sky
maps of the temperature fluctuations sourced by the kinetic SZ
effect. Maps are typically constructed by stacking outputs of the
simulated volume over a range of timesteps, often rotating or
translating each output to prevent the repetition of structures along
the line of sight and generate a larger simulated sky area.

Comparing our model predictions with those made from synthetic sky
maps is a non-trivial task. The simulations encompass a range of
cosmological parameters and often have a more limited redshift range
than our fiducial model (where $z_{\rm rei} = 10$; for simplicity we
refer to the highest redshift output of each simulation as `$z_{\rm
  rei}$'). Furthermore, as we described in detail in Section
\ref{sec:results}, the limited size of the simulation volume results
in the truncation of large-scale velocity modes and thus a significant
underestimate of the kSZ signal. For example, we estimate that a
simulation box of side length $100$ Mpc/$h$ would underestimate the kSZ
power at $\ell = 3000$ by 60\%. A box size of at least $1$ Gpc/$h$ is
required to fully account (to within 1\% of the total power) for all
the velocity modes that contribute to the kSZ power spectrum at $\ell
= 3000$.

To compare our model with that of previous studies, we must correct
their results to account for the limited simulation box size, the
variations in the maximum redshift assumed, and differing cosmological
parameters. We use our NR model as a baseline to rescale the
simulation results to our fiducial cosmology (namely, $\sigma_8 =
0.82$, $\Omega_b = 0.047$ and $\zrei = 10$), and to estimate the
amount by which the kSZ amplitude must be increased to account for
velocity mode-truncation. We also include a helium correction such
that the level of helium ionization is consistent with our fiducial
model (i.e., neutral at all epochs).

The predictions for the kSZ effect from simulations presented in
previous work are shown in Table \ref{tab:previous_simulations}. The
table shows three columns for the kSZ power at $\ell = 3000$
($\dthree$), the first shows the raw prediction for $\dthree$ taken from
each work. The second column shows the kSZ power rescaled to our
fiducial cosmology using the scalings given in Table
\ref{tab:cosmscaling_NR}. The third column shows $\dthree$ having
additionally corrected for the simulation box size in each case.

We consider the results from the smoothed-particle-hydrodynamics (SPH)
simulations of \citet[][henceforth, WHS02]{white02}, the Eulerian ENZO
simulations of \citet[][H09]{hallman09} and both the non-radiative
(NR) simulations and those including cooling, star-formation and
active galactic nucleus (AGN) feedback performed by
\citet[][B10]{battaglia10}. The WHS02 simulation includes radiative
cooling, star-formation and galactic winds (although they note these
processes may be inhibited by the limited mass resolution of their
simulation). The H09 simulation is run in the non-radiative regime. We
also include the `adiabatic' and `standard' semi-analytic models of
\citet[TBO11]{trac11} \citep[see also ][]{sehgal10}. The adiabatic
model assumes that gas resides in hydrostatic equilibrium in the
potential well of dark matter halos identified in their N-body
simulation. The standard model also includes simple prescriptions for
star-formation and non-gravitational energy feedback from supernovae
and AGN.

The cosmological rescaling for the WHS02 and H09 simulations are
fairly large as both these simulations were run assuming $\sigma_8 =
0.9$ and $z_{\rm rei} = 19$ and 3, respectively. The box-size
corrections are largest for WHS02, who used a $100$ Mpc/$h$ box, and
B10, who generated their kSZ predictions using simulation box sizes of
$165$ Mpc/$h$.

In general, the simulations predict between 2.27 and 3.91 $\mksq$,
depending on the level of gas physics included, although H09 measure
an amplitude of more than twice the upper end of this range. They are
therefore consistent with the predictions of our NR ($3.2 \mksq$) and
CSF ($2.2 \mksq$) models. B10 find that the inclusion of radiative
cooling, star-formation, and AGN feedback reduces the kSZ power by $1
\mksq$. The semi-analytic models of TBO11 lie on the low end of the
range of simulations discussed here. Their `adiabatic' model is
similar to our NR model, while their `standard' model includes
prescriptions for star-formation and energy feedback. The effect of
these prescriptions is to reduce the kSZ signal by $0.4 \mksq$. They
also explore the kSZ signal obtained when the stellar mass in groups
and clusters was increased beyond that of the standard model, finding
a further reduction of $0.3 \mksq$. This supports our assertion that
the CSF model -- which predicts the lowest amplitude of all --
provides a robust lower limit on the kSZ power spectrum due to the
reduction of the mean gas density in groups and clusters via
star-formation.

\begin{table}[ht!]
\begin{center}
\caption{\label{tab:previous_simulations} kSZ Predictions from Simulations}
\small
\begin{tabular}{c|c|c|c}
\hline\hline
\rule[-2mm]{0mm}{6mm}
Paper &   \multicolumn{3}{c}{$\dthree \;\;[\mksq]$} \\
 & Unnorm. & Cosm Cor. & Box Cor. \\
\hline
WHS02 & 3.00 & 1.57 & 3.91 \\
H09 & 7.40 & 8.50 & 9.45 \\ 
B10, NR & 2.50 & 2.42 & 4.03 \\ 
B10, AGN & 1.50 & 1.45 & 2.42 \\ 
TBO11, adiabatic & 2.50 & 2.70 & 2.70 \\ 
TBO11, standard & 2.10 & 2.27 & 2.27 \\ 
This work, DM & \nodata & \nodata & 3.44 \\
This work, NR & \nodata & \nodata & 3.24 \\
This work, CSF & \nodata & \nodata & 2.19 \\
\hline
\end{tabular}
\tablecomments{The amplitude of the kSZ power predicted by
  hydrodynamical simulations in previous work. We show the results
  from the SPH simulations of \citet[][WHS02]{white02}, the ENZO
  simulations of \citet[][H09]{hallman09}, both the non-radiative (NR)
  simulations and those including cooling, star-formation and AGN
  feedback from \citet[][B10]{battaglia10} and the `adiabatic' and
  `standard' models from the N-body plus semi-analytic approach of
  \citet[TBO11]{trac11}. }
\end{center}
\end{table}

\subsection{Comparisons with Observations}

To date, the kinetic SZ effect has not yet been detected
observationally. However, both the Atacama Cosmology Telescope and the
South Pole Telescope have placed upper limits on the amplitude of the
kSZ power at $\ell = 3000$. From 296 deg$^2$ of data,
\citet{dunkley10} obtained a $2\sigma$ upper limit of $D_\ell = 8
\mksq$.  More recently \citet{shirokoff11} obtained an upper limit of
$D_\ell = 6.5 \mksq$ (also $2\sigma$) from 210 deg$^2$. Therefore,
observations do not currently constrain any of the models that we have
presented in this work, although the predictions of the simulations of
\citet{hallman09} are inconsistent with the most recent measurements.

The main difficulty in measuring kSZ power lies with disentangling the
signal from both the thermal SZ effect and bright foregrounds,
primarily dusty, star-forming galaxies (DSFG). Unfortunately, the
angular shape of both the tSZ and DSFG power spectra is similar to
that of the kSZ signal, so it is difficult to separate these signals
in $\ell$-space. However, each has a very different frequency
dependence; multifrequency observations will be extremely effective in
separating these components. Combining SPT or ACT data with Planck and
Herschel should enable a significant detection of kSZ power in the
near future.

\section{Discussion \& Conclusion}
\label{sec:discussion}

The kinetic SZ (kSZ) power spectrum is generated by the coupling
between large scale velocity flows and small-scale density
perturbations. To predict its amplitude and shape it is necessary to
understand the behavior of the power spectrum of gas density
fluctuations over the range of scales corresponding to $0.1 \leq k
\leq 10 \hmpc$. In this work, we have introduced a new model for the
kinetic SZ power spectrum that accounts for the effect of baryonic
physics on the power spectrum of gas density fluctuations and thus on
the kSZ power spectrum.

To this end, we defined a window function, $W^2(k) = P_{\rm
  gas}(k)/P_{\rm DM}(k)$, to provide a mapping between dark matter and
gas density power spectra in our calculations. We utilized
hydrodynamic simulations -- run in both the non-radiative regime and
including radiative cooling and star-formation -- to measure the
window functions and investigated their effect on the kSZ power
spectrum. We have presented three models for the kSZ power spectrum:
DM (dark matter), in which gas density fluctuations follow those of
the dark matter at all scales, NR (non-radiative) and CSF (cooling \&
star-formation) in which we use the window functions measured in our
hydrodynamic simulations.

There is only a small difference between the DM and NR models. Gas
density fluctuations in our non-radiative simulations are suppressed
at spatial scales smaller than those that contribute significantly to
the kSZ power spectrum (for $\ell < 10,000$). At $\ell = 3000$, the NR
model predicts $\dell = 3.24 \mksq$, only $0.20 \mksq$ below the DM
model. However, the CSF power spectrum has a significantly lower
amplitude as well as a flatter shape than either the NR or DM
models. At $\ell = 3000$, the CSF model predicts $\dell = 2.19\mksq$,
$1.25 \mksq$ below the DM model. The reduction in power is driven by
the decrease in the mean gas density in group- and cluster-mass halos
due to the high-levels of cooling and star-formation in our
simulation. This in turn reduces the amplitude of the gas density
power spectrum and thus the kSZ power spectrum. 

To investigate the impact on the shape and amplitude of the kSZ power
spectrum of variations in star-formation history, we rerun our CSF
simulation having turned off radiative cooling at $z = 1.8$. We find
that the resulting window function smoothly evolved from a CSF-like
shape to a NR-like shape, with the amplitude (at large scales / low
$k$) lying midway between the two cases. The resulting kSZ power
spectrum was thus also almost directly between that of the NR and CSF
models.

We have argued that our NR and CSF models provide reasonable upper and
lower limits to the effect of astrophysical processes on the kSZ power
spectrum. The NR model is calibrated from a simulation that included
no cooling and star-formation and thus no means of reducing the gas
density. On the other hand, our CSF model is calibrated to a
simulation that suffers from the over-cooling problem, i.e., from
excessive cooling and star-formation. Hence, we expect our CSF model
to underestimate the true kSZ signal. Nevertheless, the difference
between the NR and CSF models at $\ell = 3000$ is only $1
\mksq$. Taking the mean of these two models, we find that the
astrophysical uncertainty on the kSZ power spectrum amplitude is
roughly $\pm 20\%$. This is significantly less than that of the
thermal SZ power spectrum, for which the current theoretical
uncertainty on the amplitude is $\pm 50\%$ \citep{shaw10, trac11}.

There are two caveats to this argument. Firstly, our fiducial models
assume helium remains neutral at all epochs. Singly or doubly ionized
helium would increase the amplitude of our models by up to 33\%,
depending on the redshifts at which helium reionization
occurs. Secondly, the amplitude of the kSZ power spectrum is sensitive
to cosmological parameters, namely $\sigma_8$, $\Omega_b$, $H_0$ and
$\tau$ (and, equivalently, $z_{\rm rei}$). We have investigated the
scaling of our models with cosmological parameters, finding that
$\dthree \propto \sigma_8^{4.5}\Omega_b^{2.1}H_0^{1.7} \tau^{0.44}$ for
the CSF model. The current $1\sigma$ uncertainty on $\sigma_8$
obtained from cluster number counts \citep{vikhlinin09} translates to
a $\pm 15$\% uncertainty on $\dthree$.

We have compared our models with predictions made directly from
hydrodynamic simulations \citep{white02, hallman09, battaglia10,
  trac11}.  These studies measured the kSZ power spectrum using
synthetic kSZ maps constructed by the stacking of simulation box
outputs. A serious drawback of this approach is that velocity modes
are truncated at the scale of the simulation box. For boxes of side
length $\ll 1$ Gpc/$h$, this can result in a significant underestimate
of the kSZ signal. We used our model to correct the predictions of the
simulations in previous studies for this effect, as well as
differences in cosmological parameters. We then found a reasonably
good agreement among different simulations results and the range
predicted by our models.

In this work we consider only the homogeneous- or post-reionization
contribution to the kinetic SZ signal. It is well known that models of
{\em inhomogeneous} reionization, in which different regions of the
Universe are reionized at different times, predict a {\em patchy}
kinetic SZ signal. \citet{zahn05} and \citet{mcquinn05} demonstrate
that, to first order, the magnitude of the patchy signal is dependent
on the duration of reionization. Given current technology, the kinetic
SZ power spectrum thus provides a unique probe of the redshift range
spanned by the epoch of reionization.  However, current observations
only provide a measure of the sum of the patchy and post-reionization
kSZ power spectra at angular scales around $\ell = 3000$. Therefore,
in order to extract the patchy component, it is important to have a
good theoretical understanding of the post-reionization
contribution. In this work we have performed a detailed investigation
of the theoretical uncertainty on the post-reionization kSZ power
spectrum due to astrophysical processes, cosmological parameters and
helium reionization. Adding these in quadrature produces a total
uncertainty of $\sim$30\% on the amplitude of the post-reionization
kSZ power spectrum at $\ell = 3000$.

Improvements in our understanding of the kSZ power spectrum can be
made both theoretically and observationally.  Models and simulations
of the kinetic SZ power spectrum can be further developed to make
improved theoretical predictions.  The over-cooling problem may be
mitigated by the inclusion of energy feedback from active galactic
nuclei in our simulations \citep{sijacki07,booth09, battaglia10,
  mccarthy10, teyssier11}. Feedback from AGN can heat gas sufficiently
in high density regions to slow the local cooling (and thus
star-formation) rate. \citet{daSilva01} demonstrated that the
inclusion of a preheating model to their simulations suppressed
star-formation and increased the kSZ power with respect to simulations
without preheating. On the other hand \citet{battaglia10} also
demonstrate that AGN also heat gas out to large cluster radii,
flattening the density profile, reducing the gas fraction in groups
and clusters. The net impact of AGN feedback on the SZ power spectrum
will depend on the extent to which these two effects balance one
another.  Recent models of AGN in simulations have typically been
tuned to match local ($z \sim 0$) observations, however, as we have
demonstrated, a significant contribution of the kSZ comes from larger
redshifts where the star formation histories of halos are relatively
unconstrained.  Furthermore, higher (sub-kpc) resolution simulations
may be required to resolve star-formation in lower mass halos.

From an observational perspective, small-scale CMB experiments have
currently only placed upper limits on the amplitude of the kSZ power
spectrum and do not constrain any of our models \citep{shirokoff11,
  dunkley10}. However, with increasing area and frequency coverage,
observations will become significantly more constraining in the near
future. kSZ tomography has the potential to constrain models of the
post-reionization signal by measuring its redshift dependence. This
can be achieved by cross-correlating a kSZ estimator constructed from
spectroscopic galaxy catalogues with a CMB map \citep{ho09, shao11}.
We have investigated the contribution of slices in redshift to the
total kinetic SZ power for each of our models. We found that the
principle difference between the NR and CSF models comes from sources
in the range $0 \leq z \leq 2$. Upcoming surveys have the potential to
measure the redshift distribution of the kSZ effect out to $z \sim 1$
using kSZ tomography. This therefore provides a potentially powerful
means of constraining our kSZ power spectrum models.

\section{ACKNOWLEDGMENTS}
We would like to thank Ryan Keisler, Christian Reichardt, Oliver Zhan,
Gil Holder, Wayne Hu, Nikhil Padmanabhan, Pengjie Zhang and Zheng Zheng for useful
and informative discussions.  This work was supported in part by the
NSF grant AST-1009811, by NASA ATP grant NNX11AE07G, by the Yale
University, and by the facilities and staff of the Yale University
Faculty of Arts and Sciences High Performance Computing Center.
D.H.R. acknowledges the support of NSF grant OCI-0904484.  The
simulations used in this study were performed on the SGI Altix system
(columbia) at NASA Ames and the BulldogM cluster at the Yale
University Faculty of Arts and Sciences High Performance Computing
Center.

\end{document}